\documentclass[reprint, aps, pre, notitlepage]{revtex4-2}

\usepackage{amsmath, amsfonts, amssymb, amsthm, mathtools}
\usepackage[utf8]{inputenc}
\usepackage[T1]{fontenc}
\usepackage{hyperref}
\usepackage{graphicx}
\usepackage[caption=false]{subfig}
\usepackage{lipsum}
\usepackage{xcolor}
\usepackage{empheq}

\theoremstyle{remark}

\theoremstyle{remark}

\theoremstyle{theorem}

\theoremstyle{theorem}

\begin{document}
	
\title{The one-dimensional telegraphic process with noninstantaneous stochastic resetting}
\author{Mattia Radice}
\email[Corresponding author: ]{m.radice1@uninsubria.it}
\affiliation{Dipartimento di Scienza e Alta Tecnologia and Center for Nonlinear and Complex Systems, Università degli studi dell'Insubria, Via Valleggio 11, 22100 Como, Italy}
\affiliation{I.N.F.N. Sezione di Milano, Via Celoria 16, 20133 Milano, Italy}
\begin{abstract}
In this paper we consider the one-dimensional dynamical evolution of a particle traveling at constant speed and performing, at a given rate, random reversals of the velocity direction. The particle is subject to stochastic resetting, meaning that at random times it is forced to return to the starting point. Here we consider a return mechanism governed by a deterministic law of motion, so that the time cost required to return is correlated to the position occupied at the time of the reset. We show that in such conditions the process reaches a stationary state which, for some kinds of deterministic return dynamics, is independent of the return phase. Furthermore, we investigate the first-passage properties of the system and provide explicit formulas for the mean first-hitting time. Our findings are supported by numerical simulations.
\end{abstract}
	
	\maketitle
\section{Introduction}
Stochastic processes under resetting have attracted growing interest in the last few years. Such processes may be represented by a diffusing particle which, at random or fixed times, is reset to a given position, which usually coincides with the starting point \cite{EvaMajSch-2020}. One can find models of this kind in many fields: For example in biophysics this is a good way to reproduce the Michaelis-Menten reaction scheme \cite{ReuUrbKla-2014,RotReuUrb-2015}. In biology, stochastic resetting is used to describe the competition between RNA polymerization and polymerases backtracking along the DNA template \cite{RolLisSanGri-2016}. In fields such as computer science \cite{LubSinZuc-1993,MonZec-2002} and biological physics \cite{RobHadUrb-2019}, algorithms and patterns with random restarts are identified as optimal search strategies.

In the literature there is a rich variety of examples where one analyzes the effects of resetting on different kinds of stochastic motion, ranging from Continuous-Time Random Walks \cite{MonVil-2013,MenCam-2016,Shk-2017,MonMasVil-2017,BodSok-2020-CTRWRes} to Lévy flights \cite{KusMajSabSch-2014,KusGod-2015}. Also, the domain of waiting time distributions between resetting events considered so far covers exponential distributions \cite{EvaMaj-2011}, power-laws \cite{EulMet-2016,NagGup-2016,BodSok-2020-CTRWRes} and includes the case of resetting at fixed times, which has been recognized as the most effective resetting protocol for search processes \cite{PalKunEva-2016,PalReu-2017,CheSok-2018}. Nevertheless, in the most common formulation the dynamics is governed by Brownian motion, the resetting is stochastic, occurring at constant rate \cite{EvaMaj-2011,EvaMaj-2014}, and the return is instantaneous. This picture, however, presents at least two flaws. First, Brownian motion is not always the best model to describe the stochastic dynamics of the system, especially when it is probed at microscopic scales and one needs to take into account the effects of a finite speed of propagation. A well-known case of study is the run-and-tumble motion of \textit{Escherichia coli} bacteria \cite{Ber,TaiCat-2008}, but we can find such situations when dealing with long-chain polymers \cite{Pat-1953}, chemotaxis \cite{Sch-1993} and active matter \cite{Det-2014,EscTorBroLin-2018}. In this regard, the simplest models to incorporate the momentum dynamics into the stochastic evolution of the displacement are the persistent random walk \cite{RH} and its continuous-time counterpart, known as the Telegraphic process \cite{Wei-2002}. This observation has led to an increasing interest in the study of the Telegraphic process undergoing stochastic resetting \cite{EvaMaj-2018,Mas-2019,SanBasSab-2020}. The second problem is related to the fact that it is not always possible to neglect the time cost needed to return to the initial location after the reset, for example in experiments. Indeed, in the last few years several studies have described and analyzed the experimental realization of systems where different kinds of resetting protocols are implemented with the action of optical traps \cite{BesBovPet-2020,TalPalSek-2020,MerBoyMaj-2020,SanDasNat-2021,GupPlaKun-2021}. By exerting a confining potential on the system, which may be switched on and off, an active trap forces a diffusing particle to move towards the minimum of the potential, which usually corresponds to the resetting location. In order to take into account the contribution of the return dynamics to the whole process, models including random refractory times preceding the return have been presented in the literature \cite{MasCamMen-2019,MasCamMen-2019JSTAT,EvaMaj-2018Refractory}, but these do not consider spatiotemporal correlations. For this reason, recent works introduced the idea that the return to the starting location should be performed according to a deterministic law \cite{PalKusReu-2019,PalKusReu-2019PRE,MasCamMen-2019PRE,PalKusReu-2020,BodSok-2020-BrResI}. In particular, Ref. \cite{BodSok-2020-BrResI} provides a theory to treat analytically a stochastic process subject to resetting for which the return to the starting point  is ruled by a general equation of motion.

The aim of this paper is to investigate the properties of the one-dimensional Telegraphic process undergoing stochastic Poissonian resetting with a deterministic dynamics towards the starting site, by using the analytical tools presented in \cite{BodSok-2020-BrResI}. Since explicit results can be obtained if we can determine the return time from the equation of motion, we will consider three types of return motion for which this is doable: (i) motion at constant speed; (ii) motion at constant acceleration; (iii) harmonic motion.

The paper is organized as follows: In Sec. \ref{s:resetting_rev} and Sec. \ref{s:TP_rev} we provide the reader with a short review of the theory presented in Ref \cite{BodSok-2020-BrResI} and an introduction to the Telegraphic process. In Sec. \ref{s:TP_and_R} we apply the theory to our model, presenting the results obtained for the probability density function, the stationary distribution and the corresponding moments. In Sec. \ref{s:MFPT} we investigate the first-passage properties of the system and their dependence on the return dynamics, by providing explicit expressions for the mean first-passage time. Finally, in Sec. \ref{s:conc} we draw our conclusions and discuss the results. 

\section{Model: Noninstantaneous resetting}\label{s:resetting_rev}
In this section we shortly review the model presented in \cite{BodSok-2020-BrResI} in order to provide the reader with the analytical tools that will be used throughout the paper. The process as a whole can be seen as a sequence of subprocesses. Each subprocess can be split in two phases: the displacement phase $ x(t) $, consisting in the stochastic dynamics of the diffusing particle, and the return phase, which occurs after the resetting event and consists in the deterministic motion to the resetting location. In the following we will always take $ x=0 $ as both the starting point of the displacement phase and the resetting location. The duration $ \tau $ of the displacement phase, which corresponds to the time of the resetting event measured from the start of the subprocess, is a random variable drawn from a waiting time distribution with density $ \psi(\tau) $. The deterministic motion during the return phase is governed by the equation of motion $ x=\chi(x_0,\theta) $, which can be inverted to obtain the time needed to travel from $ x_0 $ to $ x $, i.e., $ \theta=\vartheta(x_0,x) $. The time cost to return to the origin is $ \theta(x_0)=\vartheta(x_0,0) $. The total duration of the subprocess is thus $ t =\tau+\theta(x_0)$.

\subsection{Duration of a subprocess}
Let $ \phi(t) $ be the probability density that the duration of the subprocess is $ t $. The duration $ \tau $ of the displacement phase is ruled by the waiting time distribution for the resetting event $ \psi(\tau) $. At the time of the reset the particle occupies a random position $ x $, with distribution $ p(x,\tau) $, where $ p(x,t)  $ is the probability density of the displacement phase. The duration $ \theta(x_0) $ of the return phase is a deterministic function of this random position, hence we can write:
\begin{equation}\label{key}
	\phi(t)=\int_{0}^{\infty}\mathrm{d}\tau\psi(\tau)\int_{-\infty}^{+\infty}\mathrm{d}x\delta\left[t-\tau-\theta(x)\right]p(x,\tau),
\end{equation}
where $ \delta(\cdot) $ denotes the Dirac delta function. We can eliminate the constraint of the delta function by defining the Laplace transform
\begin{equation}\label{key}
	\hat{\phi}(s)=\int_{0}^{\infty}e^{-st}\phi(t)\mathrm{d}t,
\end{equation}
and performing the integration in $ t $:
\begin{equation}\label{eq:phi_LT}
	\hat{\phi}(s)=\int_{-\infty}^{+\infty}\mathrm{d}xe^{-s\theta(x)}\int_{0}^{\infty}\mathrm{d}\tau e^{-s\tau}\psi(\tau)p(x,\tau).
\end{equation}
The probability that the $ n $-th subprocess ends at time $ t $ is
\begin{equation}\label{key}
	\phi_n(t)=\int_{0}^{t}\phi_{n-1}(t')\phi(t-t')\mathrm{d}t',
\end{equation}
which can be equivalently written in Laplace space as:
\begin{equation}\label{key}
	\hat{\phi}_n(s)=\left[\hat{\phi}(s)\right]^n.
\end{equation}

The total probability of starting a new subprocess at time $ t $ is defined by
\begin{equation}\label{key}
	\kappa(t)=\delta(t)+\sum_{n=1}^{\infty}\phi_n(t),
\end{equation}
where the delta function accounts for the fact that the first subprocess starts at $ t=0 $.
Note that in the Laplace domain this becomes
\begin{equation}\label{eq:kappa_LT}
	\hat{\kappa}(s)=\frac{1}{1-\hat{\phi}(s)},
\end{equation}
hence for distribution possessing a well-defined first moment $ \eta $ such that in the small-$ s $ limit we can write $ \hat{\phi}(s)\sim1-\eta s $, this equation yields $ \hat{\kappa}(s)\sim 1/(\eta s) $, where $ \eta $ is the mean duration of a subprocess. This means that in the long-time limit the renewal rate converges to a constant value:
\begin{equation}\label{key}
	\lim_{t\to\infty}\kappa(t)=\frac{1}{\eta}.
\end{equation}

\subsection{General form of the probability density function and stationary distribution}
In order to obtain the probability density function (PDF) of the position for the complete process, $ P(x,t) $, one can first define $ G(x,t) $, the density of the displacement for a subprocess. Then $ P(x,t) $ can be expressed in terms of the probability that the last subprocess started at time $ t'<t $ and has reached $ x $ in a time $ t-t' $:
\begin{equation}\label{eq:P_complete}
	P(x,t)=\int_{0}^{t}\kappa(t')G(x,t-t')\mathrm{d}t'.
\end{equation}
Now we note that $ G(x,t) $ is the sum of two terms: Indeed, if no reset has occurred up to time $ t $ from the beginning of the subprocess, then the displacement density is simply given by $ p(x,t) $ and hence the corresponding contribution can be written as
\begin{equation}\label{eq:Q_1}
	G_1(x,t)=p(x,t)\Psi(t),
\end{equation}
where $ \Psi(t) $ denotes the probability that no resetting event has occurred up to time $ t $:
\begin{equation}\label{key}
	\Psi(t)=\int_{t}^{\infty}\psi(t)\mathrm{d}t.
\end{equation}
If instead the reset occurred at time $ \tau<t $, then the process stopped at a random position $ x_0 $, whose distribution is $ p(x_0,\tau) $, and then started moving according to the deterministic law of motion $ x=\chi\left(x_0,\theta\right) $. Therefore the corresponding term can be expressed as
\begin{equation}
	G_2(x,t)=\int_{0}^{\infty}\mathrm{d}\tau\psi(\tau)\int_{-\infty}^{+\infty}\mathrm{d}x_0p(x_0,\tau)g_2(x,t;x_0,\tau),\label{eq:Q_2}
\end{equation}
where $g_2(x,t;x_0,\tau)$ is
\begin{multline}\label{key}
	g_2(x,t;x_0,\tau) = \delta \left[x-\chi(x_0,t-\tau)\right]\times\\
	\Theta(t-\tau)\Theta(\tau+\theta(x_0)-t),
\end{multline}
and $ \Theta(x) $ is defined as follows:
\begin{equation}\label{key}
	\Theta(x)=\begin{cases}
		1 & \text{if }x\geq 0\\
		0 & \text{if }x<0.
	\end{cases}
\end{equation}

In order to evaluate the PDF of the complete process, we observe that since Eq. \eqref{eq:P_complete} takes the form of a convolution, it is more easily evaluated in Laplace space:
\begin{equation}\label{eq:P_complete_LT}
	\hat{P}(x,s)=\hat{\kappa}(s)\hat{G}(x,s)=\frac{\hat{G}(x,s)}{1-\hat{\phi}(s)},
\end{equation}
where we used Eq. \eqref{eq:kappa_LT} and we recall that $ \hat{G}(x,s)=\hat{G}_1(x,s)+\hat{G}_2(x,s) $. By using Eqs. \eqref{eq:Q_1} and \eqref{eq:Q_2} it is possible to obtain more explicit formulas for the Laplace transforms $ \hat{G}_1(x,s) $ and $ \hat{G}_2(x,s) $. For $ \hat{G}_1(x,s) $ we get:
\begin{equation}\label{eq:G1_LT}
	\hat{G}_1(x,s)=\int_{0}^{\infty}e^{-st}p(x,t)\Psi(t)\mathrm{d}t.
\end{equation}
For $ \hat{G}_2(x,s) $, we change the order of integration between $ t $ and $ \tau $, and by introducing $ u= t-\tau $ we can write:
\begin{align}
	\hat{G}_2(x,s)=&\int_{0}^{\infty}\mathrm{d}\tau e^{-s\tau}\psi(\tau)\int_{-\infty}^{+\infty}\mathrm{d}x_0 p(x_0,\tau)\times\nonumber\\
	&\int_{-\tau}^{\infty}\mathrm{d}u e^{-su}\delta\left[x-\chi(x_0,u)\right]\Theta(u)\Theta(\theta(x_0)-u)\\
	=&\int_{0}^{\infty}\mathrm{d}\tau e^{-s\tau}\psi(\tau)\int_{-\infty}^{+\infty}\mathrm{d}x_0 p(x_0,\tau)\times\nonumber\\
	&\int_{0}^{\infty}\mathrm{d}u e^{-su}\delta\left[x-\chi(x_0,u)\right]\Theta(\theta(x_0)-u)\label{eq:G2_interm},
\end{align}
We now observe that the integration variable $ u $ represents the time of the return phase, during which the distance of the particle from the origin is a decreasing function of $ u $. This means that the delta function in the last integral provides a nonvanishing contribution only for distances smaller than the initial distance, i.e., for $ |x|\leq|x_0| $. The computation can then be performed by using the equation of motion to define a new integration variable $ y $ as $ y=\chi(x_0,u) $. This definition can be inverted to write the old variable $ u $ as
\begin{equation}\label{key}
	u=\vartheta(x_0,y),
\end{equation}
so that the integral in $ u $ yields:
\begin{multline}\label{key}
	\int_{0}^{\infty}\mathrm{d}u e^{-su}\delta\left[x-\chi(x_0,u)\right]\Theta(\theta(x_0)-u)=\\
	\begin{dcases}
		\frac{e^{-s\vartheta(x_0,x)}}{\lvert v(x_0,x)\rvert} & \text{for }|x|\leq|x_0|\\
		0 & \text{for }|x|>|x_0|,
	\end{dcases}
\end{multline}
where $ |v(x_0,x)|=\left\lvert\left(\tfrac{\mathrm{d}\vartheta(x_0,x)}{\mathrm{d}x} \right)^{-1}\right\rvert$ is the speed of the particle at position $ x $. Therefore we find:
\begin{multline}\label{eq:G2_LT}
	\hat{G}_2(x,s)=\int_{0}^{\infty}\mathrm{d}\tau e^{-s\tau}\psi(\tau)\int_{|x|}^{\infty}\mathrm{d}x_0\times\\
	\begin{dcases}
		\frac{e^{-s\vartheta(x_0,x)}}{|v(x_0,x)|}p(x_0,\tau)&\text{for }x\geq0\\
		\frac{e^{-s\vartheta(-x_0,x)}}{|v(-x_0,x)|}p(-x_0,\tau)&\text{for }x<0,
	\end{dcases}
\end{multline}
and one can prove that the resulting PDF is correctly normalized, see Appendix \ref{app:norm}.

The large-$ t $ behaviour of $ P(x,t) $ can be deduced from the small-$ s $ limit of $ \hat{P}(x,s) $. The situation is particularly easy if the mean duration of a subprocess is finite. Indeed, from Eq. \eqref{eq:P_complete_LT}, by taking the limit $ s\to 0 $ and assuming that $ \hat{\phi}(s)\sim 1-\eta s $, we can write
\begin{equation}\label{key}
	\hat{P}(x,s)\sim \frac{1}{\eta s}\hat{G}(x,s=0)
	=\frac{1}{\eta s}\int_{0}^{\infty}G(x,t)\mathrm{d}t,
\end{equation}
from which one can deduce the following relation in the time domain:
\begin{equation}\label{key}
	\lim_{t\to\infty}P(x,t)=P(x)=\frac 1\eta\left[\rho_1(x)+\rho_2(x)\right],
\end{equation}
where
\begin{align}\label{key}
	\rho_1(x)&=\int_{0}^{\infty}G_1(x,t)\mathrm{d}t\\
	\rho_2(x)&=\int_{0}^{\infty}G_2(x,t)\mathrm{d}t.
\end{align}
This means that in the long-time limit the system reaches a steady state, often referred to as \emph{nonequilibrium steady state} (NESS) for to the presence of persistent probability currents directed towards the resetting position \cite{EvaMaj-2014}. The steady state is described by $ P(x) $, the \emph{stationary distribution}, which in this case is the sum of two different contributions. The components $ \rho_1(x) $ and $ \rho_2(x) $ can be computed as the limit $ s\to 0 $ of $ \hat{G}_1(x,s) $ and $ \hat{G}_2(x,s) $, respectively, yielding the expressions:
\begin{equation}\label{eq:rho1}
		\rho_1(x)=\int_{0}^{\infty}p(x,t)\Psi(t)\mathrm{d}t,
\end{equation}
and
\begin{multline}\label{eq:rho2}
	\rho_2(x)=\int_{0}^{\infty}\mathrm{d}\tau\psi(\tau)\int_{|x|}^{\infty}\mathrm{d}x_0\times\\
	\begin{dcases}
		\frac{p(x_0,\tau)}{|v(x_0,x)|}&\text{for }x\geq0\\
		\frac{p(-x_0,\tau)}{|v(-x_0,x)|}&\text{for }x<0.
	\end{dcases}
\end{multline}

We point out that there exist cases in which, despite the action of a resetting mechanism, the system never reaches a steady state. This fact has been observed, e.g., for Scaled Brownian Motion under Poissonian and power-law resetting \cite{BodCheSok-2019}, or in diffusive systems with resetting at power-law times \cite{EulMet-2016,NagGup-2016,PalKunEva-2016}. In such situations,  the long-time behavior of the system can still be described by an invariant density which is, however, non-normalized. In other words, for some exponent $ \alpha $ there exists
\begin{equation}\label{key}
	\lim_{t\to\infty} t^\alpha P(x,t)=\mathcal{I}_{\infty}(x),
\end{equation}
where $ \mathcal{I}_{\infty}(x) $  is called an \emph{infinite density} in the sense that it is non-normalizable. Nevertheless, the infinite density can capture some of the statistical properties of the system. The study of infinite invariant densities and the connection with the properties of stochastic processes is a major topic in infinite ergodic theory, which has attracted significant interest in recent years, see for example \cite{KesBar-2010,DecLutBar-2011,VezBarBur-2019,Lei-Bar,ROAP,AghKesBar-2020,BarRadAki-2021,Far-PRE}.

\section{The Telegraphic process}\label{s:TP_rev}
The Telegraphic process has been widely considered in both the physical and the mathematical literature \cite{Kac-1974,Ors-1990,MasPorWei-1992,Ors-1995,Wei-2002,GarOrsPol-2014}. In one dimension the process may be defined as the mathematical equivalent of a particle moving at constant velocity $ c $ and randomly reversing the direction of motion from time to time at a fixed rate $ \gamma $. If $ c_0 $ is the initial velocity, then the state of the velocity at time $ t $ is
\begin{equation}\label{key}
	v(t)=c_0(-1)^{N(t)},
\end{equation}
where $ N(t) $ is the number of events up to time $ t $ of a homogeneous Poisson process of rate $ \gamma $. It follows the Kac's representation \cite{Kac-1974}:
\begin{equation}\label{eq:TP_repr}
	x(t)=\int_{0}^{t}v(t')\mathrm{d}t'=c_0\int_{0}^{t}(-1)^{N(t')}\mathrm{d}t'.
\end{equation}
It turns out that the equation for the time evolution of the PDF of the process is the so-called Telegrapher's equation \cite{Wei-2002}
\begin{equation}\label{eq:TE}
	\frac{\partial ^2 p}{\partial t^2}+2\gamma\frac{\partial p}{\partial t}=c^2\frac{\partial^2 p}{\partial x^2},
\end{equation}
whose solution, relative to the initial conditions
\begin{equation}\label{eq:TE_initial}
	p(x,0)=\delta(x),\quad \left.\frac{\partial p(x,t)}{\partial t}\right\rvert_{t=0}=0,
\end{equation}
is \cite{Wei-2002}:
\begin{multline}\label{eq:TE_sol}
	p(x,t)=\frac{e^{-\gamma t}}{2}\bigg\lbrace\delta(x-ct)+\delta(x+ct)+\\
	\frac{\gamma}{c}\left[I_0(z)+\frac{\gamma tI_1(z)}{z}\right]\Theta(ct-|x|)\bigg\rbrace,
\end{multline}
where $ I_0(z) $ and $ I_1(z) $ are modified Bessel functions of the first kind \cite{Abr-Steg} and $ z $ is the dimensionless variable defined as
\begin{equation}\label{key}
	z=\frac{\gamma}{c}\sqrt{c^2t^2-x^2}.
\end{equation}
Contrarily to the diffusion equation, the Telegrapher's equation contains the speed of the particle $ c $ as a parameter, implying that the probability density propagates in space with finite velocity. Indeed, the solution presented in Eq. \eqref{eq:TE_sol} vanishes unless $ -ct\leq x\leq ct $. However, note that we can still recover the diffusion equation if we take the limit $ c\to\infty $, $ \gamma\to\infty $, with the ratio
\begin{equation}\label{key}
	\frac{c^2}{2\gamma}=D
\end{equation}
kept constant (\textit{diffusive limit}). Indeed, one can show that the mean square displacement is \cite{Wei-2002}
\begin{equation}\label{key}
	\langle x^2(t)\rangle = \frac{c^2}{2\gamma^2}\left(2\gamma t-1+e^{-2\gamma t}\right),
\end{equation}
which grows linearly in time and reduces to $ \langle x^2(t)\rangle =2Dt $ in the aforementioned limit. On the other hand, the limit $ \gamma\to0 $ corresponds to a wave equation
\begin{equation}\label{key}
	\frac{\partial^2 p}{\partial t^2}=c^2\frac{\partial^2 p}{\partial x^2},
\end{equation}
describing the deterministic motion at fixed $ c $ of a particle never reversing the velocity direction (\textit{ballistic limit}).

In the following we will need the Laplace transform of $ p(x,t) $, which can be obtained by first computing the probability density in Fourier-Laplace space. By defining
\begin{equation}\label{key}
	\tilde{p}(k,s)=\int_{0}^{\infty}\mathrm{d}te^{-st}\int_{-\infty}^{+\infty}\mathrm{d}ke^{-ikx}p(x,t),
\end{equation}
plugging this expression into the Telegrapher's equation and considering the initial conditions specified by Eq. \eqref{eq:TE_initial}, one obtains
\begin{equation}\label{key}
	\tilde{p}(k,s)=\frac{s+2\gamma}{s(s+2\gamma)+c^2k^2},
\end{equation}
whose Fourier inversion is easy to perform, since $ \tilde{p}(k,s) $ presents two simple imaginary poles:
\begin{equation}\label{key}
	k_\pm = \pm\frac ic\sqrt{s(s+2\gamma)}.
\end{equation}
Therefore, one finally gets:
\begin{equation}\label{eq:TE_sol_LT}
	\hat{p}(x,s)=\int_{-\infty}^{+\infty}\frac{\mathrm{d}k}{2\pi}e^{ikx}\tilde{p}(k,s)=\frac{\lambda_s}{2s}e^{-|x|\lambda_s},
\end{equation}
where the parameter $ \lambda_{q} $ is defined as:
\begin{equation}\label{eq:lambda_def}
	\lambda_q=\frac{\sqrt{q(q+2\gamma)}}{c}.
\end{equation}

\section{Telegraphic process and Poissonian resetting}\label{s:TP_and_R}
We now consider the Telegraphic process undergoing stochastic resetting and performing the return motion at constant velocity, constant acceleration and under the effect of a harmonic potential. The particle starts from $ x(0)=0 $ randomly choosing the initial direction of  motion, with equal probability. The position evolves according to Eq. \eqref{eq:TP_repr}. The dynamics is interrupted at a random time $ \tau $ by the resetting event, after which the particle starts moving towards the origin according to the deterministic equation of motion. When the origin is finally reached, the process is restored to the initial conditions and starts anew. Note that different resetting protocols for the initial state of the velocity can also be considered \cite{EvaMaj-2018}. We will take Poissonian resetting, so that the distribution of the time interval between resetting events is exponential
\begin{equation}\label{eq:psi_exp}
	\psi(\tau)=re^{-r\tau},
\end{equation}
where $ r $ is the resetting rate. It immediately follows that the probability of no resetting occurring in the time interval $ (0,t) $ is:
\begin{equation}\label{Psi_exp}
	\Psi(t)=r\int_{t}^{\infty}e^{-r\tau}\mathrm{d}\tau=e^{-rt}.
\end{equation}

\subsection{Return motion at constant speed}
\begin{figure}
	\includegraphics[width=8.6cm]{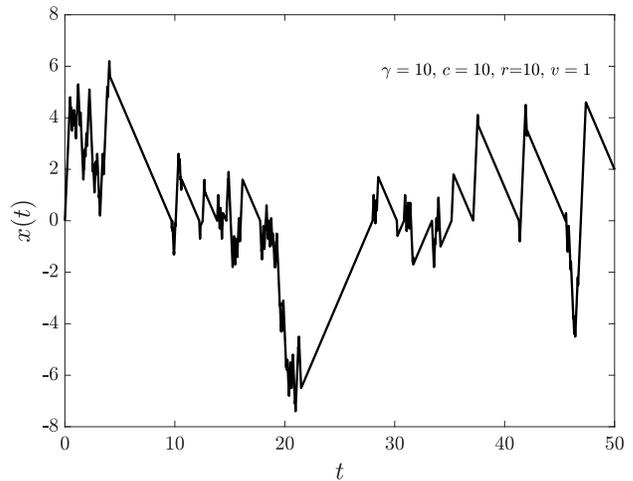}
	\caption{A sample trajectory of the Telegraphic process undergoing stochastic Poissonian resetting and returns at finite velocity.}
	\label{fig:Traj_vconst}
\end{figure}

Let $ v $ be the absolute value of the velocity. Then if the particle is reset at position $ x_0 $, its position during the return motion evolves according to
\begin{equation}\label{eq:motion_v}
	x(\theta,x_0)=-\mathrm{sgn}(x_0)\cdot v\theta+x_0,
\end{equation}
where $ \mathrm{sgn}(\cdot) $ denotes the sign of the argument. The time cost  needed to return to the origin is thus:
\begin{equation}\label{eq:theta_constant_v}
	\theta(x_0)=\frac{|x_0|}{v}.
\end{equation}
A sample trajectory is represented in Fig. \ref{fig:Traj_vconst}.

The Laplace transform of the duration of a subprocess can be computed from Eq. \eqref{eq:phi_LT} by considering the return time given by Eq. \eqref{eq:theta_constant_v}, see Appendix \ref{app:mean_duration}. We obtain
\begin{equation}\label{eq:phi_v}
	\hat{\phi}(s)=\frac{r}{r+s}\cdot\frac{v\lambda_{r+s}}{s+v\lambda_{r+s}},
\end{equation}
with $ \lambda_{r+s} $ defined by Eq. \eqref{eq:lambda_def}. The corresponding mean duration $ \eta $ is:
\begin{equation}\label{key}
	\eta=\frac 1r+\frac 1{v\lambda_r}.
\end{equation}
Note that since we are considering Poissonian resetting, the term $ 1/r $ is the mean duration of the displacement phase, while the remaining term is the contribution of the return motion.

We now proceed with the computation of the PDF in Laplace space. The contribution of the displacement phase is given by Eq. \eqref{eq:G1_LT}, from which we see that, in virtue of Eq. \eqref{Psi_exp}, the function $ \hat{G}_1(x,s) $ is expressed as the Laplace transform of $ p(x,t) $, with $ r+s $ playing the role of the Laplace variable. We thus have
\begin{equation}\label{eq:G1_expression_v}
	\hat{G}_1(x,s)=\hat{p}(x,r+s)=\frac{\lambda_{r+s}}{2(r+s)}e^{-|x|\lambda_{r+s}},
\end{equation}
see Eq. \eqref {eq:TE_sol_LT}. The other contribution can instead be written, for $ x\geq0 $, as
\begin{align}
	\hat{G}_2(x,s)&=\frac{1}{v}\int_{x}^{\infty}\mathrm{d}x_0e^{-s\vartheta(x_0,x)}\int_{0}^{\infty}\mathrm{d}\tau e^{-s\tau}\psi(\tau)p(x_0,\tau)\\
	&=\frac rv\int_{x}^{\infty}\mathrm{d}x_0e^{-s\vartheta(x_0,x)}\int_{0}^{\infty}\mathrm{d}\tau e^{-(r+s)\tau}p(x_0,\tau)\\
	&=\frac rv\int_{x}^{\infty}\mathrm{d}x_0e^{-s\vartheta(x_0,x)}\hat{p}(x_0,r+s),
\end{align}
where $ \vartheta(x_0,x) $ can be deduced from the equation of motion Eq. \eqref{eq:motion_v}:
\begin{equation}\label{key}
	\vartheta(x_0,x)=
	\begin{dcases}
		\frac{x_0-x}{v}&\text{for }x\geq0\\
		\frac{x-x_0}{v}&\text{for }x<0.
	\end{dcases}
\end{equation}
By using this expression in the former integral, we get
\begin{equation}\label{eq:G2_expression_v}
	\hat{G}_2(x,s)=\frac{r}{s+v\lambda_{r+s}}\hat{p}(x,r+s),\quad x\geq0,
\end{equation}
and one can check that the same result is valid for the case $ x<0 $. By adding the contributions given by Eqs. \eqref{eq:G1_expression_v} and \eqref{eq:G2_expression_v} we obtain the Laplace transform $ \hat{G}(x,s) $, which can be plugged in Eq. \eqref{eq:P_complete_LT}, together with the expression for $ \hat{\phi}(s) $ given by Eq. \eqref{eq:phi_v}. This yields
\begin{equation}\label{key}
	\hat{P}(x,s)=\frac{r+s}{s}\hat{p}(x,r+s)=\frac{\lambda_{r+s}}{2s}e^{-|x|\lambda_{r+s}},
\end{equation}
which can be inverted explicitly, so that one arrives at:
\begin{equation}\label{eq:PDF_v_t}
	P(x,t)=e^{-rt}p(x,t)+r\int_{0}^{t}e^{-rt'}p(x,t')\mathrm{d}t'.
\end{equation}
Interestingly, this is the same density one obtains in the case of Poissonian resetting with instantaneous returns, see for instance \cite{EvaMaj-2018,EvaMajSch-2020}. Therefore in this case the PDF is completely unaffected by the particular type of return dynamics. Furthermore, by taking the limit $ t\to\infty $, we can check that the system relaxes to the steady state
\begin{equation}\label{eq:stationaryP_av}
	 P(x)=\lim_{t\to\infty}P(x,t) =\frac{\lambda_r}{2}e^{-|x|\lambda_r},
\end{equation}
which indeed is the same obtained for instantaneous resetting \cite{EvaMaj-2018,Mas-2019}.
One can compute the $ q $-th moment of the stationary distribution, obtaining
\begin{equation}\label{eq:mom_av}
	\langle|x|^q\rangle=\lambda_r\int_0^{\infty}x^qe^{-\lambda_r x}\mathrm{d}x=\frac{\Gamma(1+q)}{\lambda_r^q},
\end{equation}
and in particular the mean square displacement is:
\begin{equation}\label{key}
	\langle x^2\rangle=\frac{2}{\lambda_r^2}=\frac{2c^2}{r(r+2\gamma)}.
\end{equation}

We remark that the independence of the steady state from the return velocity was already observed in the literature for many systems, including the Telegraphic process \cite{PalKusReu-2019}. Such a feature may be extended to the finite-time behavior of the PDF, as it happens in the case of Brownian motion \cite{PalKusReu-2019PRE}.

\subsection{Return motion at constant acceleration}
\begin{figure}
 	\includegraphics[width=8.6cm]{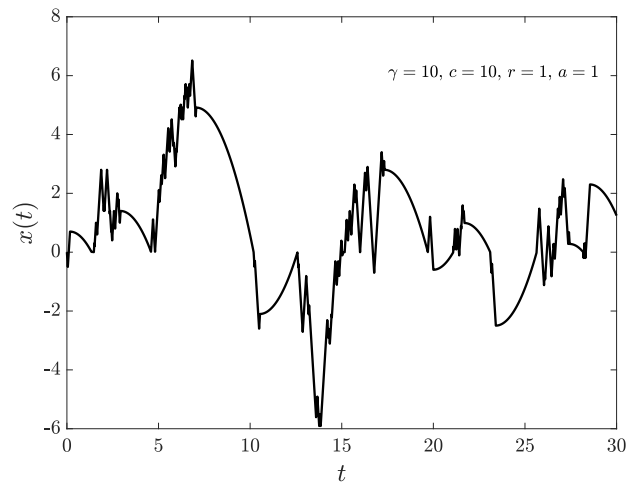}
 	\caption{A sample trajectory of the Telegraphic process undergoing stochastic Poissonian resetting and returns at finite acceleration.}
 	\label{fig:Traj_aconst}
 \end{figure}

We assume that when the deterministic return motion starts, the initial velocity is $ v_0=0 $. The dynamics is thus given by
\begin{empheq}[left=\empheqlbrace]{align}\label{key}
	x(\theta,x_0)&=-\mathrm{sgn}(x_0)\cdot\frac 12a\theta^2+x_0\\
	v(\theta)&=-\mathrm{sgn}(x_0)\cdot a\theta,
\end{empheq}
where $ a $ is the absolute value of the acceleration. The time needed to return to the origin from $ x_0 $ is:
\begin{equation}\label{key}
	\theta(x_0)=\sqrt{\frac{2|x_0|}{a}}.
\end{equation}
Figure \ref{fig:Traj_aconst} displays a sample trajectory.

In Appendix \ref{app:mean_duration} we show that the Laplace transform of the duration of a subprocess is \begin{equation}\label{eq:phi_a}
	\hat{\phi}(s)=\frac{r}{r+s}\left\{1-\sqrt{\pi}\xi e^{\xi^2}\left[1-\mathrm{erf}\left(\xi\right)\right]\right\},
\end{equation}
where
\begin{equation}\label{key}
	\xi=\left[\frac{2a}{cs^2}\sqrt{\left(2\gamma+r+s\right)\left(r+s\right)}\right]^{-\frac 12},
\end{equation}
and $ \mathrm{erf}(z) $ is the error function
\begin{equation}\label{key}
	\mathrm{erf}(z)=\frac{2}{\sqrt{\pi}}\int_{0}^{z}e^{-t^2}\mathrm{d}t,
\end{equation}
from which it follows that the mean duration is
\begin{equation}\label{key}
	\eta=\frac{1}{r}+\sqrt{\frac{\pi}{2a\lambda_r}}.
\end{equation}

In order to compute the PDF, we first consider the Laplace domain. The contribution of the displacement phase does not depend on the return motion, hence $ \hat{G}_1(x,s) $ has the same expression as before, given by Eq. \eqref{eq:G1_expression_v}. The contribution of the return phase instead can be computed from Eq. \eqref{eq:G2_LT}. By considering the case $ x\geq0 $ (the case $ x<0 $ follows from symmetry) and carrying out the integration in $ \tau $, we get
\begin{equation}\label{key}
	\hat{G}_2(x,s)=r\int_{x}^{\infty}\frac{\mathrm{d}x_0}{|v(x_0,x)|}e^{-s\vartheta(x_0,x)}\hat{p}(x,r+s),
\end{equation}
with
\begin{align}\label{key}
	\vartheta(x_0,x)&=\sqrt{\frac{2(x_0-x)}{a}}\\
	\left\lvert v(x_0,x)\right\rvert&=\sqrt{2a(x_0-x)}.
\end{align}
The integral thus yields
\begin{equation}\label{key}
	\hat{G}_2(x,s)=\frac rs\sqrt{\pi}\xi e^{\xi^2}\left[1-\mathrm{erf}\left(\xi\right)\right]\hat{p}(x,r+s),
\end{equation}
which is also valid for $ x<0 $. The Laplace transform $ \hat{G}(x,s) $ is then obtained by adding the two contributions $ \hat{G}_1(x,s) $ and $ \hat{G}_2(x,s) $. Therefore, from Eqs. \eqref{eq:P_complete_LT} and \eqref{eq:phi_a}, one arrives at
\begin{equation}\label{key}
	\hat{P}(x,s)=\frac{r+s}{s}\hat{p}(x,r+s),
\end{equation}
which is precisely the result of the previous case, meaning that also for returns at constant acceleration the PDF is unaffected by the return motion.


It is possible to show that, if we consider the Telegraphic process and Poissonian resetting, a sufficient condition for the independence of the PDF from the return dynamics  is that both $ \vartheta(x_0,x) $ and $ v(x_0,x) $ are functions of the distance $ |x_0-x| $:
\begin{align}\label{key}
	\vartheta(x_0,x)&=\vartheta\left(\left|x_0-x\right|\right)\\
	v(x_0,x)&=v\left(\left|x_0-x\right|\right).
\end{align}
We recall that $ \vartheta(x_0,x) $ represents the time needed to travel from $ x_0 $ to $ x $ and $ v(x_0,x) $ is the velocity reached at position $ x $ during the return phase. The two are thus related by:
\begin{equation}\label{key}
	\frac{1}{v(x_0,x)}=\frac{\partial\vartheta(x_0,x)}{\partial x}.
\end{equation}
Let us consider the case $ x\geq0 $. Then from Eq. \eqref{eq:G2_LT}, by taking $ \psi(\tau)=r\exp(-r\tau) $ and performing the time integration, we get:
\begin{align}\label{key}
	\hat{G}_2(x,s)&=r\int_{x}^{\infty}\mathrm{d}x_0\frac{e^{-s\vartheta(x_0-x)}}{|v(x_0-x)|}\hat{p}(x_0,r+s)\\
	&=r\int_{0}^{\infty}\mathrm{d}y\frac{e^{-s\vartheta(y)}}{|v(y)|}\hat{p}(y+x,r+s),
\end{align}
where in the second line we introduced the variable $ y=x_0-x $. Now, since we are considering the Telegraphic process, we have
\begin{align}\label{key}
	\hat{p}(x+y,r+s)&=\frac{\lambda_{r+s}}{2(r+s)}e^{-(x+y)\lambda_{r+s}}\\
	&=e^{-y\lambda_{r+s}}\hat{p}(x,r+s),
\end{align}
hence $ \hat{G}_2(x,s) $ can be written:
\begin{equation}\label{key}
	\hat{G}_2(x,s)=r\mathcal{C}(r,s)\hat{p}(x,r+s),
\end{equation}
where
\begin{equation}\label{key}
	\mathcal{C}(r,s)=\int_{0}^{\infty}\mathrm{d}y\frac{e^{-s\vartheta(y)-y\lambda_{r+s}}}{|v(y)|}.
\end{equation}
We remark that $ \mathcal{C}(r,s) $ depends on the parameters of the return motion. One can easily verify that the same expression is obtained for the case $ x<0 $, where it is useful to consider the change of variable $ y=x-x_0 $, so that
\begin{align}\label{key}
	\hat{G}_2(x,s)&=r\int_{-\infty}^{x}\mathrm{d}x_0\frac{e^{-s\vartheta(x-x_0)}}{|v(x-x_0)|}\hat{p}(x_0,r+s)\\
	&=r\int_{0}^{\infty}\mathrm{d}y\frac{e^{-s\vartheta(y)}}{|v(y)|}\hat{p}(x-y,r+s)\\
	&=r\int_{0}^{\infty}\mathrm{d}y\frac{e^{-s\vartheta(y)}}{|v(y)|}\hat{p}(|x|+y,r+s),
\end{align}
where in the last line we exploited the symmetry of $ \hat{p}(x,s) $. Therefore, the complete PDF reads:
\begin{align}\label{key}
	\hat{P}(x,s)&=\frac{\hat{G}_1(x,s)+\hat{G}_2(x,s)}{1-\hat{\phi}(s)}\\
	&=\frac{1+r\mathcal{C}(r,s)}{1-\hat{\phi}(s)}\hat{p}(x,r+s).
\end{align}
At this point, we observe that by integrating both sides in $ x $, the normalization condition implies
\begin{equation}\label{key}
	\frac 1s=\frac{1+r\mathcal{C}(r,s)}{1-\hat{\phi}(s)}\frac{1}{r+s},
\end{equation}
from which it immediately follows:
\begin{equation}\label{key}
	\hat{P}(x,s)=\frac{r+s}{s}\hat{p}(x,r+s).
\end{equation}
Note that the crucial point is the factorization of $ \hat{p}(x+y,r+s) $ as the product of a function of $ x $ only and a function of $ y $ only. Therefore the same reasoning can be applied to the case of Brownian motion, where
\begin{equation}\label{key}
	p(x,t)=\frac{e^{-\tfrac{x^2}{4Dt}}}{\sqrt{4\pi Dt}}\iff\hat{p}(x,s)=\frac{e^{-|x|\sqrt{\tfrac sD}}}{\sqrt{4Ds}}.
\end{equation}

\begin{figure*}
	\subfloat{
		\includegraphics[width=8.6cm]{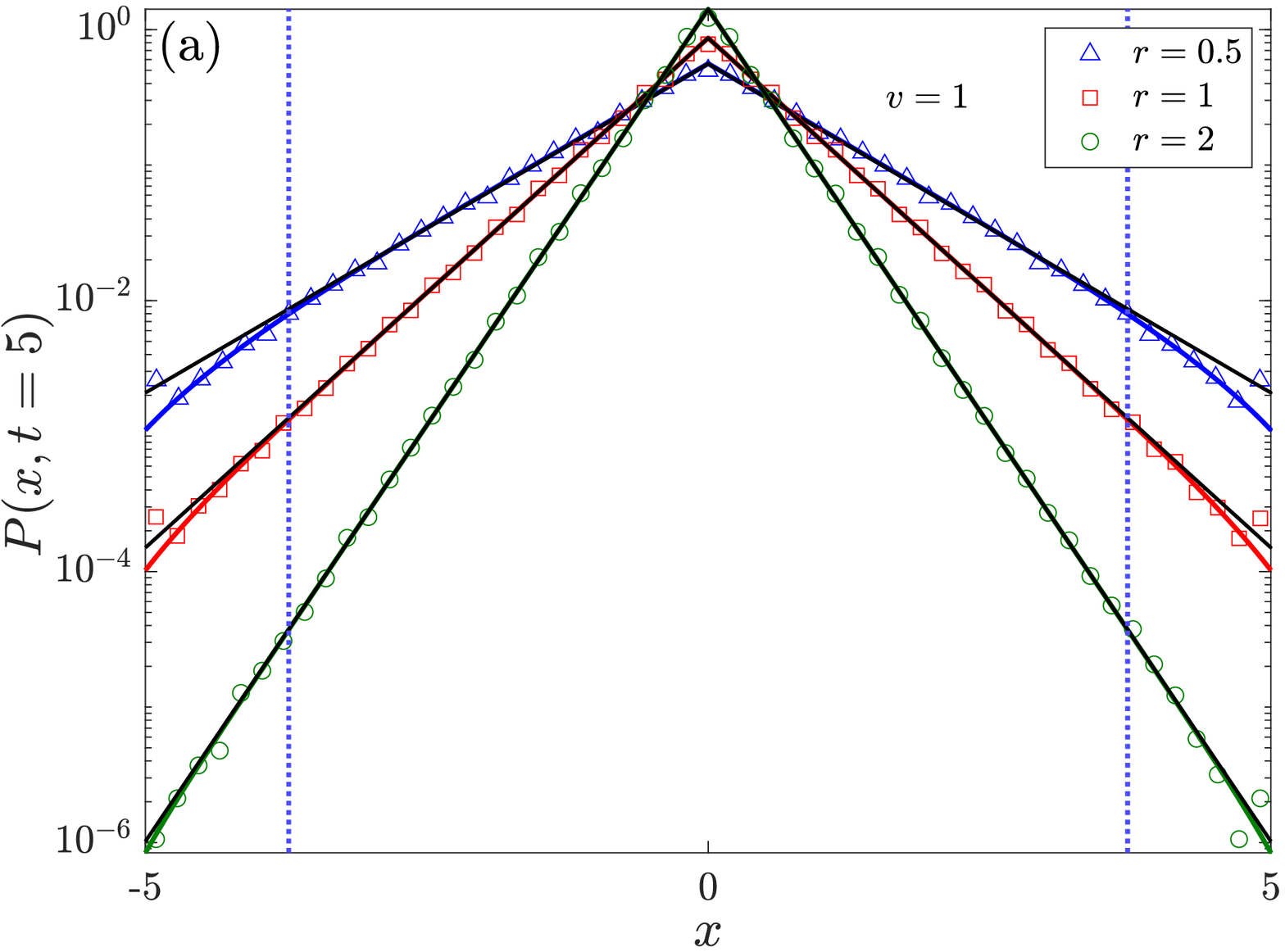}%
	}\quad
	\subfloat{
		\includegraphics[width=8.6cm]{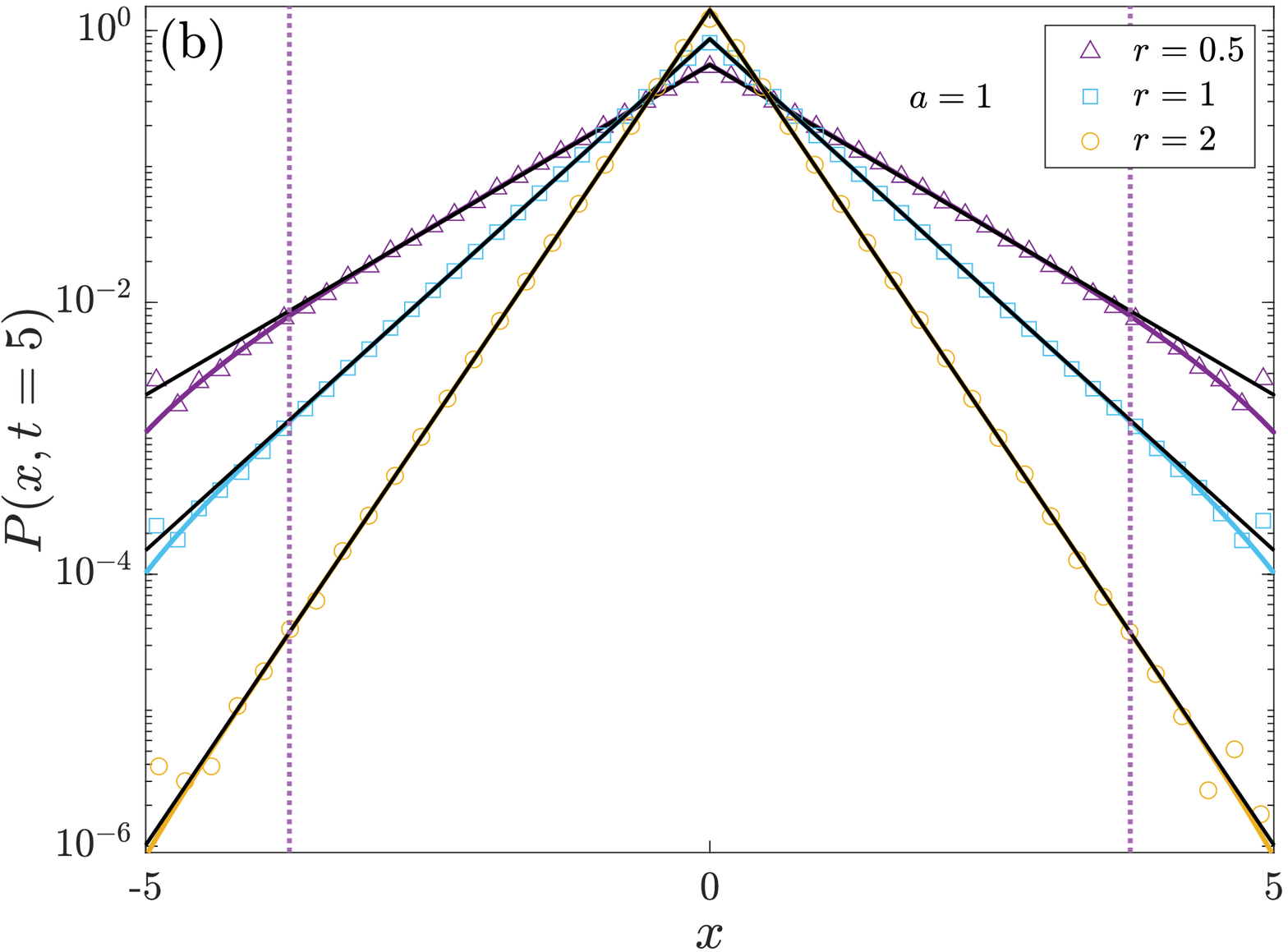}%
	}\hfill
	\caption{Agreement of the numerical results with $ P(x,t) $ given by Eq. \eqref{eq:PDF_v_t}, for several resetting rates in the case of (a) return motion at constant speed with $ v=1 $, and (b) constant acceleration with $ a=1 $. The data (markers) are compared with the theoretical time-dependent PDFs (solid lines) and the corresponding stationary distributions (black lines). In both cases we simulated $ 10^7 $ processes with individual time step length $ \mathrm{d}t=0.01 $ and set $ \gamma=c=1 $. The vertical lines mark the values $ \pm x^* $ given by Eq. \eqref{eq:x_c} in the particular case $ r=0.5 $. }
	\label{fig:PDF_Mom_avconst}
\end{figure*}

Figure \ref{fig:PDF_Mom_avconst} shows the agreement of our numerical simulation with the theoretical results. The data obtained by evolving the process numerically in the case of returns at constant speed and constant acceleration agree with the same stationary distribution, except for large values of $ |x| $. We trace back such disagreement to two main reasons: The first is the contribution of the singular part of $ p(x,t) $, i.e, the part containing the delta functions in Eq. \eqref{eq:TE_sol}, which represents those walks that do not experience any velocity reversal up to time $ t $, see \cite{Wei-2002}. This contribution is described by two ballistic peaks at $ x=\pm ct $,  whose height is proportional to the probability that a particle performs ballistic motion up to time $ t $. The ballistic peaks are relevant also in the presence of resetting, even if their probability decays in time as $ \exp\left[-(r+\gamma)t\right] $ - while for the resetting-free process, the probability of the ballistic peaks decay as $ \exp(-\gamma t) $. More details are given in Appendix \ref{s:Relax}. This fact is responsible of the deviations of the data points at the boundaries of the spatial domain, which indeed are peaked with respect to neighboring points. The second reason for the disagreement for large $ |x| $ is that, as observed for other systems \cite{MajSabSch-2015,PalKunEva-2016,Gup-2019}, the process for large but finite $ t $ relaxes to the steady state inside a spatial region delimited by a time-dependent threshold $ \pm x^* $, outside of which the system persists in a transient state, representing large deviations from the stationary distribution. By following \cite{MajSabSch-2015}, we obtain that for large $ t $ the PDF behaves as $ P(x,t)\sim\exp\left[-tI(x,t)\right] $, with
\begin{equation}\label{key}
	I(x,t)=\begin{dcases}
		\lambda_{r}\frac{|x|}{t}&\text{for }|x|<|x^*|\\
		r+\gamma-\gamma\sqrt{1-\frac{x^2}{c^2 t^2}}&\text{for }|x|>|x^*|,
	\end{dcases}
\end{equation}
and we can estimate $ |x^*| $ as
\begin{equation}\label{eq:x_c}
	|x^*|=\frac{\sqrt{r(r+2\gamma)}}{r+\gamma}\cdot ct,
\end{equation}
more detailed calculations are presented in Appendix \ref{s:Relax}. Note that for any positive $ \gamma $ the region $ (-x^*,x^*) $ is included in $ (-ct,ct) $, which is the maximum extension of the PDF for finite $ t $ (the system propagates with finite speed $ c $).

\subsection{Harmonic return motion}
\begin{figure}
	\includegraphics[width=8.6cm]{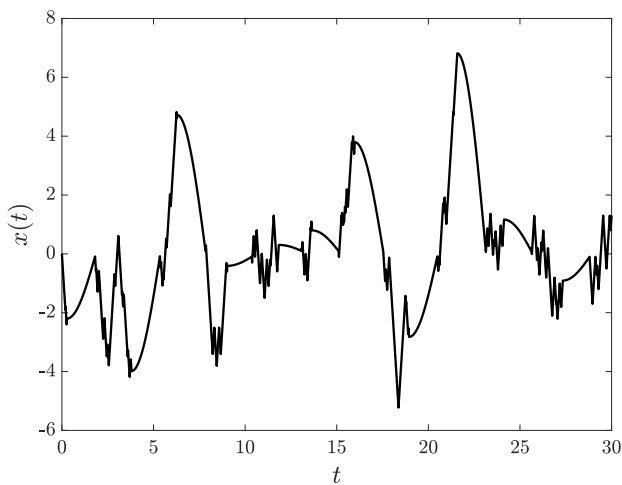}
	\caption{A sample trajectory of the Telegraphic process undergoing stochastic Poissonian resetting and harmonic return motion.}
	\label{fig:Traj_arm}
\end{figure}
A sample trajectory depicting this situation is showed in Fig. \ref{fig:Traj_arm}. In this case, by assuming that the return phase starts with $ v_0=0 $, we have the following equation of motion:
\begin{empheq}[left=\empheqlbrace]{align}
	x(x_0,\theta)&=x_0\cos\left(\omega\theta\right)\\
	v(x_0,\theta)&=-\omega x_0\sin\left(\omega\theta\right),
\end{empheq}
where $ \omega $ is the radial frequency. The time to travel to the origin is a fixed value independent of the starting point
\begin{equation}\label{key}
	\theta=\frac{\pi}{2\omega},
\end{equation}
so that the Laplace transform of the duration of a subprocess is easy to obtain, see Appendix \ref{app:mean_duration}, and can also be inverted explicitly:
\begin{align}\label{key}
	\hat{\phi}(s)&=\frac{r}{r+s}e^{-\tfrac{\pi s}{2\omega}}\\
	\phi(t)&=re^{-r\left(t-\tfrac\pi{2\omega}\right)}\Theta\left(t-\frac{\pi}{2\omega}\right).
\end{align}
The mean duration of a subprocess is trivially:
\begin{equation}\label{key}
	\eta=\frac{1}{r}+\frac{\pi}{2\omega}.
\end{equation}

The PDF can be computed following the same procedure of the previous sections. The contribution $ \hat{G}_1(x,s) $ of the displacement phase remains evidently unchanged. For the return phase, we have
\begin{align}\label{key}
	\vartheta(x_0,x)&=\frac 1\omega\arccos\left(\frac{x}{x_0}\right)\\
	\left\lvert v(x_0,\theta)\right\rvert&=\omega\sqrt{x_0^2-x^2},
\end{align}
thus the integral defining $ \hat{G}_2(x,s) $, for all $ x $, reads:
\begin{equation}\label{key}
	\hat{G}_2(x,s)=\frac r\omega\int_{|x|}^{\infty}\frac{e^{-\tfrac s\omega\arccos(|x|/x_0)}}{\sqrt{x_0^2-x^2}}\hat{p}(x_0,r+s)\mathrm{d}x_0,
\end{equation}
where we used the fact that $ \hat{p}(x_0,r+s)=\hat{p}(-x_0,r+s) $. It is convenient to consider the rescaled variable $ z=x_0/|x| $. Then, in virtue of Eq. \eqref{eq:P_complete_LT}, the complete PDF is
\begin{multline}\label{key}
	\hat{P}(x,s)=\frac{\frac 12\lambda_{r+s}}{s+r\left(1-e^{-\frac{\pi s}{2\omega}}\right)}\bigg\{e^{-|x|\lambda_{r+s}}+\\
	\frac r\omega\int_{1}^{\infty}\frac{\exp\left[-z|x|\lambda_{r+s}- s/\omega\arccos(1/z)\right]}{\sqrt{z^2-1}}\mathrm{d}z\bigg\}.
\end{multline}
In the small-$ s $ limit this expression yields
\begin{multline}\label{key}
	\hat{P}(x,s)\sim\frac1s\cdot\frac{1}{\frac 1r+\frac{\pi}{2\omega}}\bigg(\frac{\lambda_{r}}{2r}e^{-|x|\lambda_{r}}+\\
	\frac{\lambda_r}{2\omega}\int_{1}^{\infty}\frac{e^{-z|x|\lambda_r}}{\sqrt{z^2-1}}\mathrm{d}z\bigg),
\end{multline}
where the integral on the r.h.s can be written in terms of the modified Bessel function of the second kind of order $ 0 $ \cite{Abr-Steg}. The stationary distribution is thus
\begin{equation}\label{eq:stationary_arm}
	P(x)=\frac{\frac{\lambda_r}{2r}e^{-|x|\lambda_r}+\frac{\lambda_r}{2\omega}K_0\left(\lambda_r|x|\right)}{{\frac 1r+\frac{\pi}{2\omega}}},
\end{equation}
and we observe that $ P(x) $ this time depends explicitly on the parameter $ \omega $ of the return motion.
The moments of the distribution can be computed explicitly. We first note that
\begin{align}\label{key}
	\int_{0}^{\infty}x^qK_0\left(\lambda x\right)\mathrm{d}x&=\frac1{\lambda^{q+1}}\int_{0}^{\infty}\mathrm{d}z\int_{1}^{\infty}\mathrm{d}t\frac{e^{-zt}}{\sqrt{t^2-1}}\\
	&=\frac{\Gamma(1+q)}{\lambda^{q+1}}\int_{1}^{\infty}\frac{\mathrm{d}t}{t^{1+q}\sqrt{t^2-1}},
\end{align}
where $ \Gamma(\cdot) $ is the Gamma function. The last integral can be transformed by considering the change of variable $ 1/t=\cos\phi $, yielding
\begin{equation}\label{key}
	\int_{0}^{\tfrac{\pi}{2}}\cos^q\phi\mathrm{d}\phi=2^{q-1}\frac{\Gamma^2\left(\frac{1+q}{2}\right)}{\Gamma(1+q)},
\end{equation}
see Eq. 3.621(1) in \cite{GraRyz}. Therefore, the $ q $-th moment reads
\begin{equation}\label{eq:moments_arm}
	\langle|x|^q\rangle=\frac{\left[2\omega\Gamma(1+q)+r2^q\Gamma^2\left(\frac{1+q}{2}\right) \right]}{2\omega+\pi r}\lambda^{-q},
\end{equation}
and in particular the expression for the mean square displacement is:
\begin{equation}\label{key}
	\langle x^2\rangle=\frac{4\omega+\pi r}{2\omega+\pi r}\cdot\frac{c^2}{r(r+2\gamma)}.
\end{equation}

\begin{figure*}
	\subfloat{
		\includegraphics[width=8.6cm]{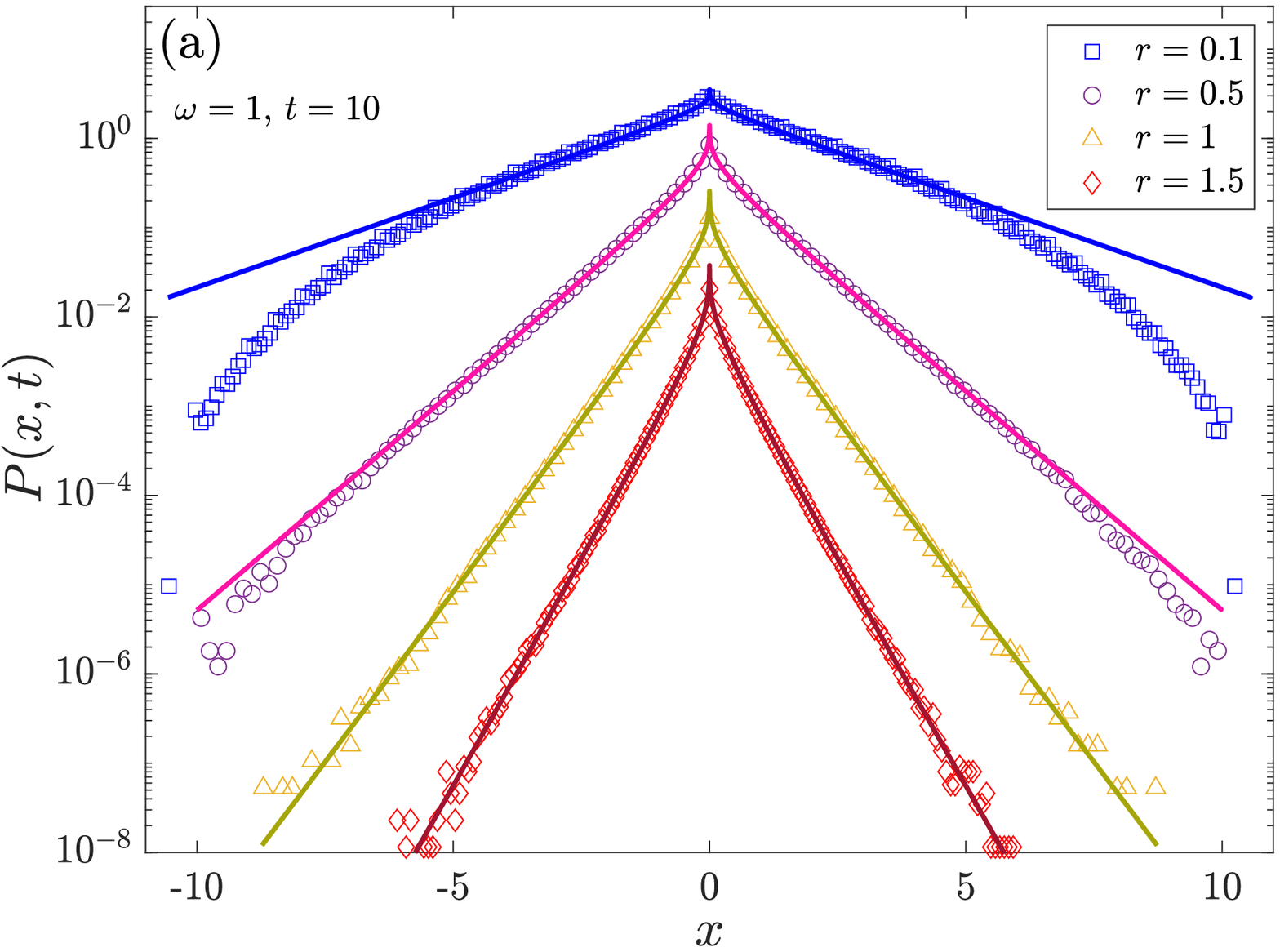}%
	}\quad
	\subfloat{
		\includegraphics[width=8.6cm]{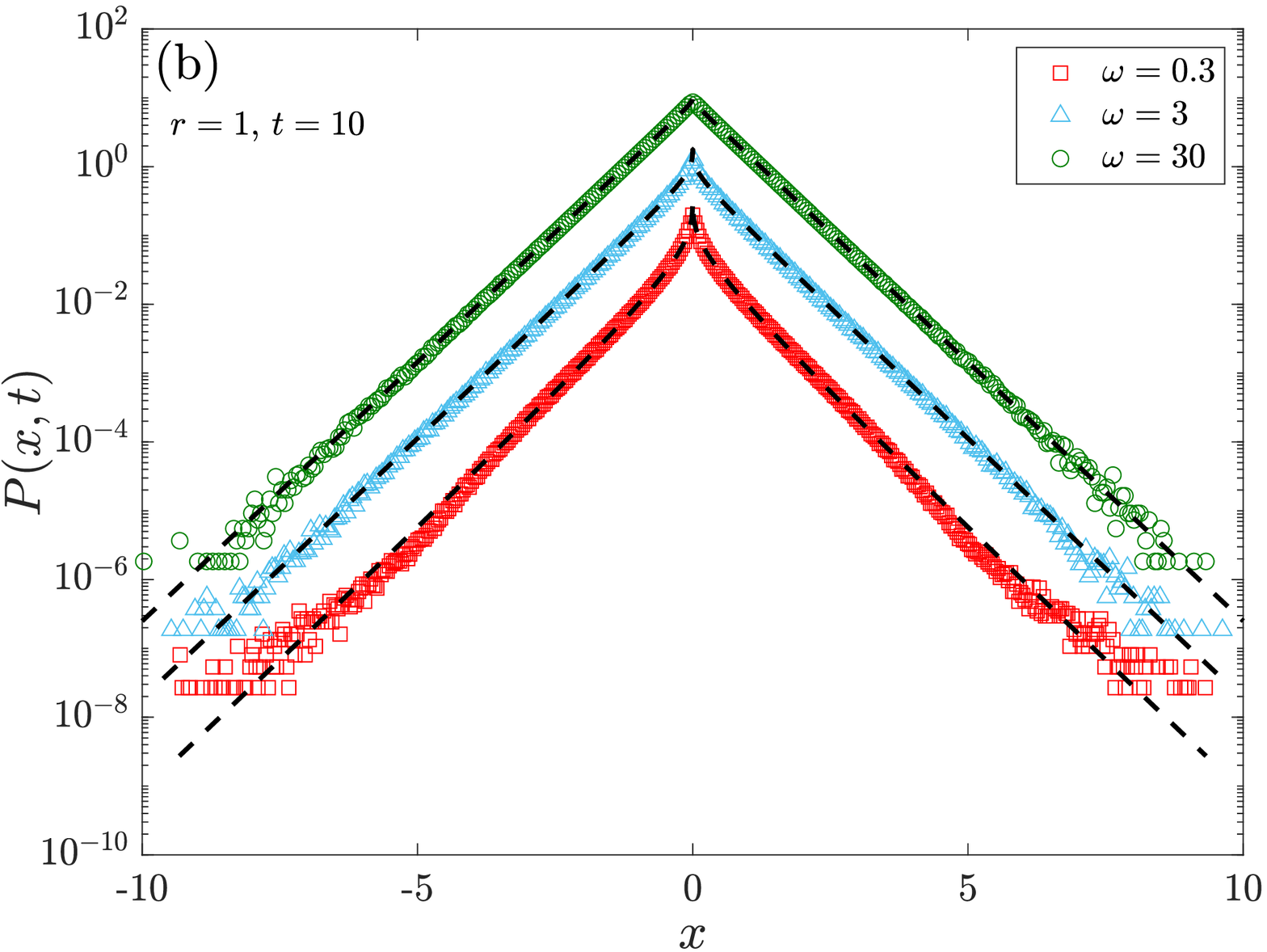}%
	}
	\caption{Agreement of the numerical results with the stationary distribution, given by Eq. \eqref{eq:stationary_arm}. (a) Distributions for fixed $ \omega $ and several $ r $. The data are the results of $ 10^7 $ simulations with step duration $ \mathrm{d}t=0.01 $. (b) Distributions for fixed $ r $ and a few $ \omega $. For each $ \omega $ we considered $ 10^8 $ processes with $ \mathrm{d}t=0.01 $. For all simulations, we took $ \gamma=c=1 $. Note that in both panels the data have been shifted to enhance the readability of the plots.}
	\label{fig:PDF_Mom_arm}
\end{figure*}

Figure \ref{fig:PDF_Mom_arm} shows the agreement of numerical simulations with the theoretical results. In particular we observe that the convergence of the numerical data to the stationary distribution is slower for lower values of the resetting rate. Indeed by the time of our simulations the data corresponding to smaller $ r $ present important deviations from $ P(x) $, especially for large $ |x| $. This suggests that the relaxation to the steady state may present similar features to those discussed in the previous cases, such as the dependence on the resetting rate and the separation of the spatial domain in a inner core region where the NESS has been established and an outer region where the system is transient. However, contrarily to the cases of returns at constant $ v $ or $ a $, the simulations also suggest that the approach to the NESS depends on the parameter $ \omega $ of the deterministic phase: The data corresponding to smaller values of $ \omega $ present more noticeable deviations with respect to $ P(x) $, in particular for large $ |x| $.

\section{Mean first-passage time}\label{s:MFPT}
Due to the constraint of a finite speed of propagation, the first-passage properties of the Telegraphic process are rather peculiar and show a much richer behavior with respect to those of Brownian motion. Let us consider a particle starting from the origin with equal probability of moving initially to the right or left, and suppose that a target is placed at position $ b>0 $. Then the first-passage time distribution reads \cite{Ors-1995,MalJemKun-2018}:
\begin{multline}\label{eq:f_Tel}
	f(b,t)=\frac{e^{-\gamma t}}{2}\delta(t-b/c)+\frac b2\cdot\frac{\gamma e^{-\gamma t}}{ct+b}\Theta(t-b/c)\times\\
	\bigg\lbrace I_0(z)+\frac{I_1(z)}{\zeta}
	\bigg[1+\frac{c\zeta^2}{\gamma b}\bigg]\bigg\rbrace,
\end{multline}
where we called
\begin{align}\label{key}
	z&=\frac\gamma c\sqrt{c^2t^2-b^2}\\
	\zeta &=\sqrt{\frac{ct-b}{ct+b}}.
\end{align}
By translating this expression in Laplace space, one obtains \cite{Ors-1995,MalJemKun-2018,EvaMaj-2018}
\begin{equation}\label{eq:f:_Tel_LT}
	\hat{f}(b,s)=\frac{1}{2\gamma}\left[s+2\gamma-\sqrt{s(s+2\gamma)}\right]e^{-b\lambda_s},
\end{equation}
where $ \lambda_s $ is defined by Eq. \eqref{eq:lambda_def}. This yields a diverging first derivative around $ s=0 $, which implies an infinite mean first-passage time, as in the case of Brownian motion. However, contrarily to Brownian motion where, independently of the initial distance, for any $ t>0 $ there is a finite probability of hitting the target, for the Telegraphic process the probability vanishes for $ t<b/c $. The standard result is recovered in the diffusive limit, i.e., by taking both $ c\to\infty $ and $ \gamma\to\infty $, with $ c^2/2\gamma=D $ kept constant. By retaining only the leading terms in the asymptotic expansion of the Bessel functions one indeed finds:
\begin{equation}\label{key}
	f(b,t)\sim\frac{b}{\sqrt{4\pi Dt^3}}e^{-\tfrac{b^2}{4Dt}}.
\end{equation}
In the opposite situation instead, i.e., in the ballistic limit $ \gamma\to 0 $, one may easily verify that only the term proportional to the delta function is nonvanishing, reflecting the fact that a wave propagating in the right direction with speed $ c $ surely hits the target at time $ t=b/c $.

In order to study the first-passage properties of the process undergoing noninstantaneous Poissonian resetting, it is particular convenient to adopt the approach of Ref. \cite{CheSok-2018}. Let us call a subprocess \textit{successful} if the particle hits the target placed at $ b>0 $ at time $ t $ during the displacement phase, and \textit{unsuccessful} if the particle is reset to the initial position before reaching the target. Let us denote with $ F(t) $ the probability density of hitting the target for the first time at time $ t $, $ \varpi(t) $ the probability density of hitting the target at time $ t $ after the start of the successful subprocess and $ \varphi(t) $ the probability density of the duration of an unsuccessful subprocess. Then $ F(t) $ is equal to the probability that the $ n $-th unsuccessful ended at time $ t'<t $ and the target is then reached in a time $ t-t' $, summed over all possible values of $ n $:
\begin{equation}\label{key}
	F(t)=\varpi(t)+\sum_{n=1}^{\infty}\int_{0}^{t}\mathrm{d}t'\varpi\left(t-t'\right)\varphi_n\left(t'\right),
\end{equation}
where $ \varphi_n(t) $ is the probability density of ending the $ n $-th unsuccessful subprocess at time $ t $. This equation can be written more conveniently in Laplace space
\begin{equation}\label{key}
	\hat{F}(s)=\hat{\varpi}(s)\sum_{n=0}^{\infty}\hat{\varphi}^n(s)=\frac{\hat{\varpi}(s)}{1-\hat{\varphi}(s)},
\end{equation}
from which one is able to obtain the mean first-passage time as:
\begin{equation}\label{eq:mean_FPT}
	\langle T\rangle=-\left.\frac{\mathrm{d}\hat{F}(s)}{\mathrm{d} s}\right\rvert_{s=0}=-\frac{\hat{\varpi}'(0)}{1-\hat{\varphi}(0)}-\frac{\hat{\varpi}(0)\hat{\varphi}'(0)}{\left[1-\hat{\varphi}(0)\right]^2}.
\end{equation}
Note that the quantities we have defined thus far also depend on the position of the target, but we drop the dependence in order to have cleaner expressions.

The probability $ \varpi(t) $ is equal to the first-passage probability of the displacement phase, which we denote as $ f(t) $, times the probability of not having been reset up to time $ t $. Thus we can write for the Laplace transform $ \hat{\varpi}(s) $
\begin{equation}\label{key}
	\hat{\varpi}(s)=\int_{0}^{\infty}e^{-st}\Psi(t)f(t)\mathrm{d}t,
\end{equation}
meaning that for Poissonian resetting, i.e., $ \Psi(t)=\exp(-rt) $, the Laplace transform is simply equal to:
\begin{equation}\label{eq:varpi_vs_f_LT}
	\hat{\varpi}(s)=\hat{f}(s+r).
\end{equation}
Hence $ \hat{\varpi}(s) $ only depends on $ \hat{f}(s) $, which is a property of the displacement phase. In the case of the Telegraphic process the function $ \hat{f}(s) $ is given by Eq. \eqref{eq:f:_Tel_LT}. The duration of an unsuccessful subprocess instead can be expressed as
\begin{equation}\label{key}
	\varphi(t)=\int_{0}^{\infty}\mathrm{d}\tau\psi(\tau)\int_{-\infty}^{b}\mathrm{d}x\delta\left[t-\tau-\theta(x)\right]q(x,\tau),
\end{equation}
where $ q(x,t) $ is the \textit{survival probability density function}, namely, the PDF of being at position $ x $ at time $ t $ and having not hit the target up to time $ t $. The corresponding Laplace transform then can be computed in the following way:
\begin{equation}\label{key}
	\hat{\varphi}(s)=\int_{-\infty}^{b}\mathrm{d}xe^{-s\theta(x)}\int_{0}^{\infty}\mathrm{d}\tau e^{-s\tau}\psi(\tau)q(x,\tau),
\end{equation}
where $ \theta(x) $ is the time needed to return to the origin from $ x $, determined by the law of the return motion. We observe that for exponential resetting, $ \psi(t)=r\exp(-rt) $, the result of the time integration can be written in terms of  $ \hat{q}(x,s) $:
\begin{equation}\label{eq:phi_LT_surv}
	\hat{\varphi}(s)=r\int_{-\infty}^{b}e^{-s\theta(x)}\hat{q}(x,s+r)\mathrm{d}x.
\end{equation}
The computation of $ q(x,t) $ for the Telegraphic process is a challenging problem, because the standard method of the images \cite{Red} used in the case of Brownian motion does not yield the correct result. This is due to finite-memory effects of the driving noise, see the discussion in the introduction of Ref. \cite{LeDMajSch}. When approaching the problem from the point of view of the Fokker-Planck equation, one has to deal with nontrivial boundary conditions \cite{Wei-1984}. The solution is considered for example in Refs. \cite{MasPorWei-1992,LeDMajSch} and we present our derivation in Appendix \ref{app:surv}, wherein we obtain:
\begin{equation}\label{eq:q_LT}
	\hat{q}(x,s)=\frac{\lambda_s}{2s}e^{-\lambda_s|x|}-\frac{\lambda_s}{2\gamma s}\left(s+\gamma-c\lambda_s\right)e^{-\lambda_s(2b-x)}.
\end{equation}
This expression can be inserted in Eq. \eqref{eq:phi_LT_surv} to compute $ \hat{\varphi}(s) $ after we specify the return motion and determine the return time $ \theta(x) $. Note that by integrating Eq. \eqref{eq:q_LT} one gets the survival probability in Laplace space:
\begin{align}\label{key}
	\hat{Q}(b,s)&=\int_{-\infty}^{b}\hat{q}(x,s)\mathrm{d}x\\
	&=\frac 1s-\frac{1}{2\gamma s}\left(s+2\gamma-c\lambda_s\right)e^{-b\lambda_s}.
\end{align}
Furthermore, by using the following relation between the survival and first passage probability $ \hat{f}(b,s) $
\begin{equation}\label{key}
	\hat{f}(b,s)=1-s\hat{Q}(b,s),
\end{equation}
one also gets the expression for $ \hat{f}(b,s) $ given by Eq. \eqref{eq:f:_Tel_LT}. In the following we present our results for the three cases of return motion previously considered.

\subsection{Return motion at constant speed}
\begin{figure*}
	\subfloat{
		\includegraphics[width=8.6cm]{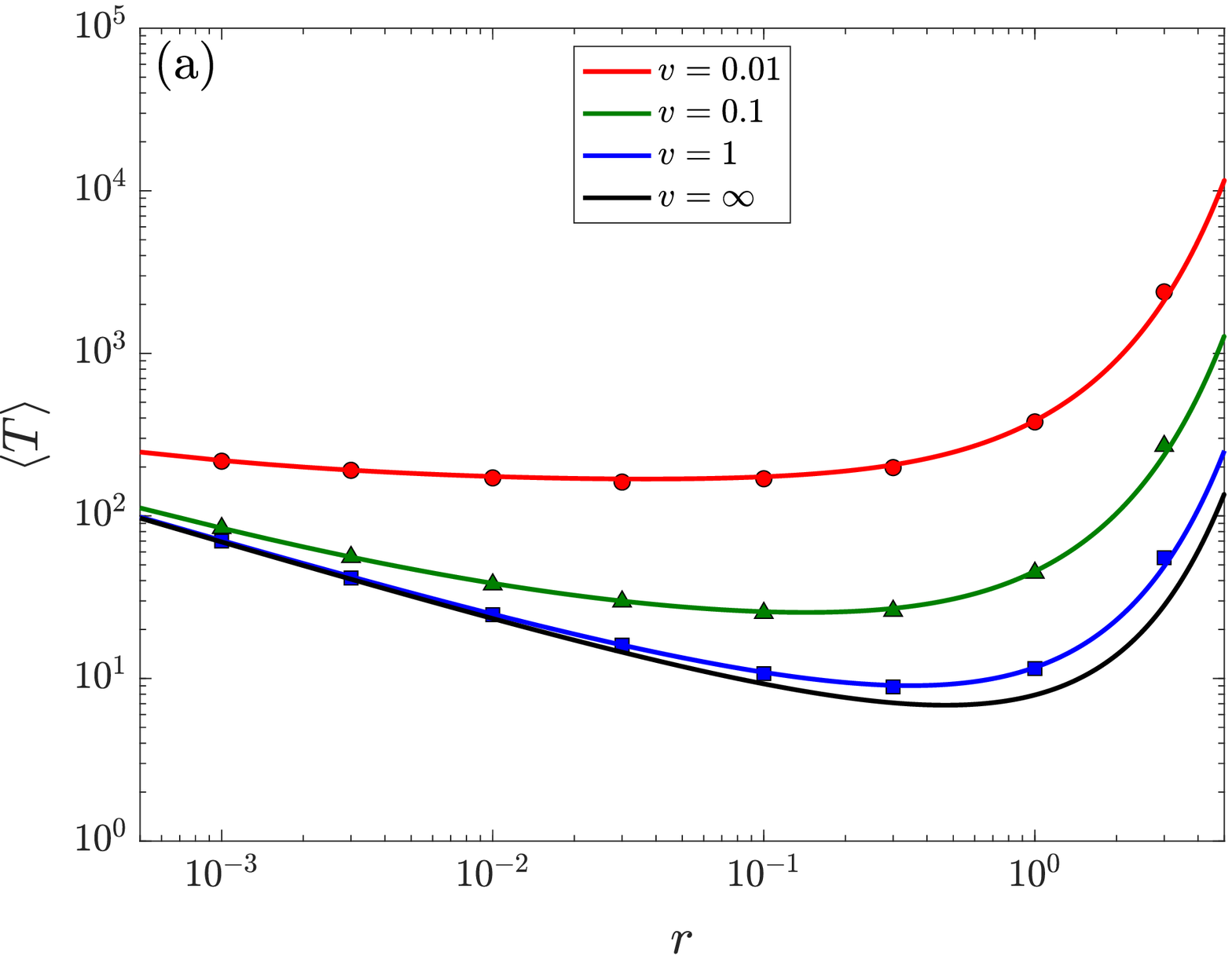}%
	}\quad
	\subfloat{
		\includegraphics[width=8.6cm]{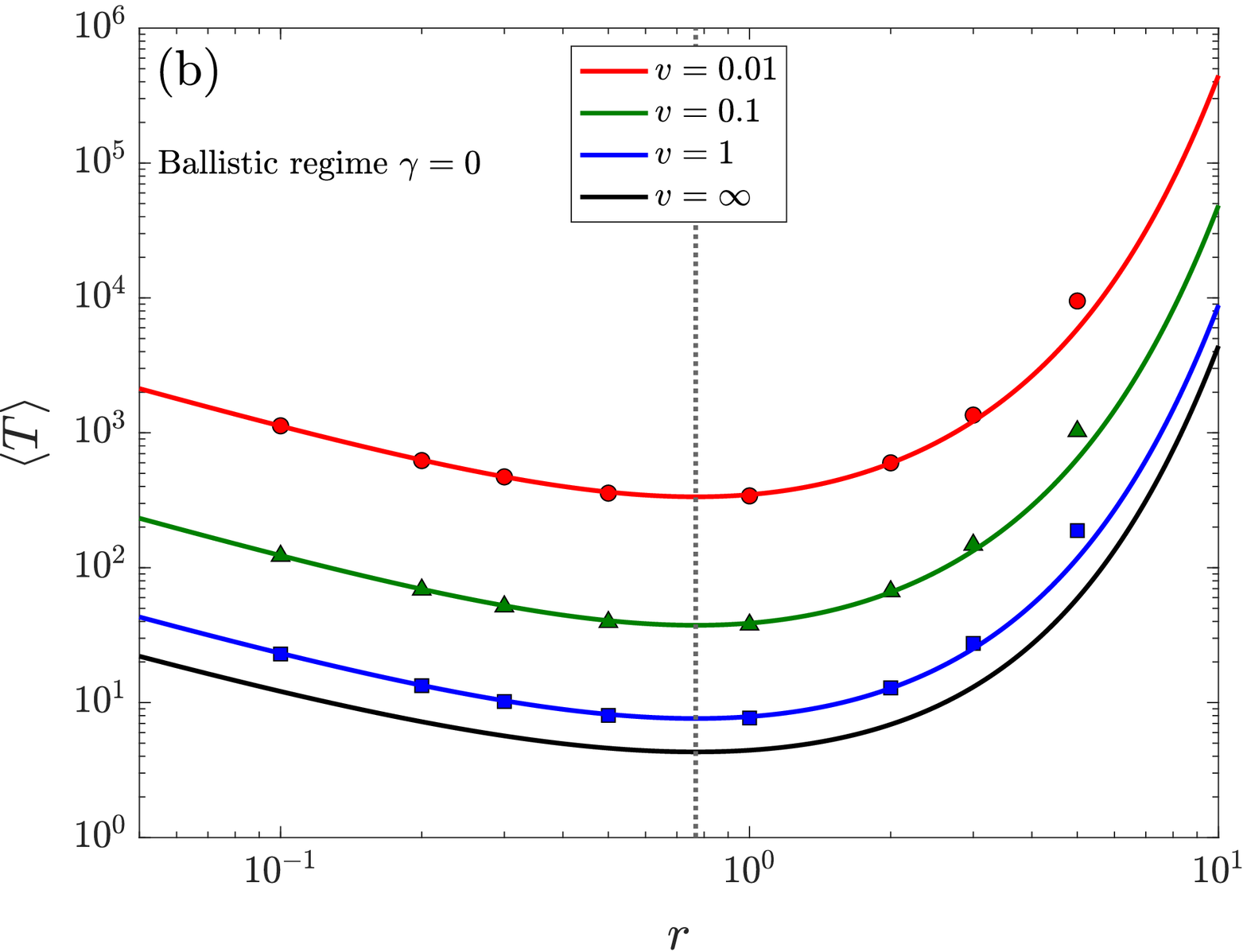}%
	}\hfill
	\subfloat{
		\includegraphics[width=8.6cm]{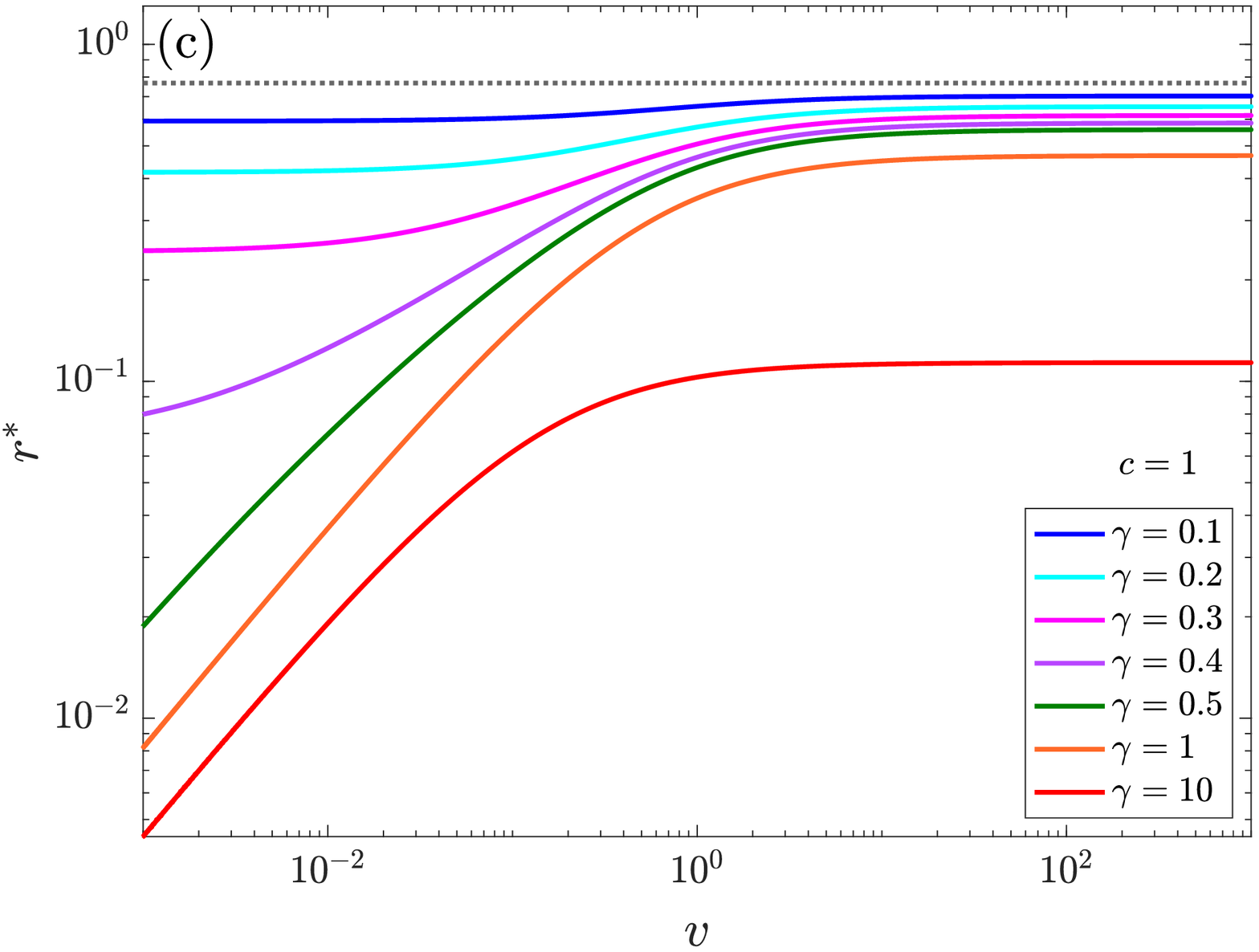}%
	}\quad
	\subfloat{
		\includegraphics[width=8.6cm]{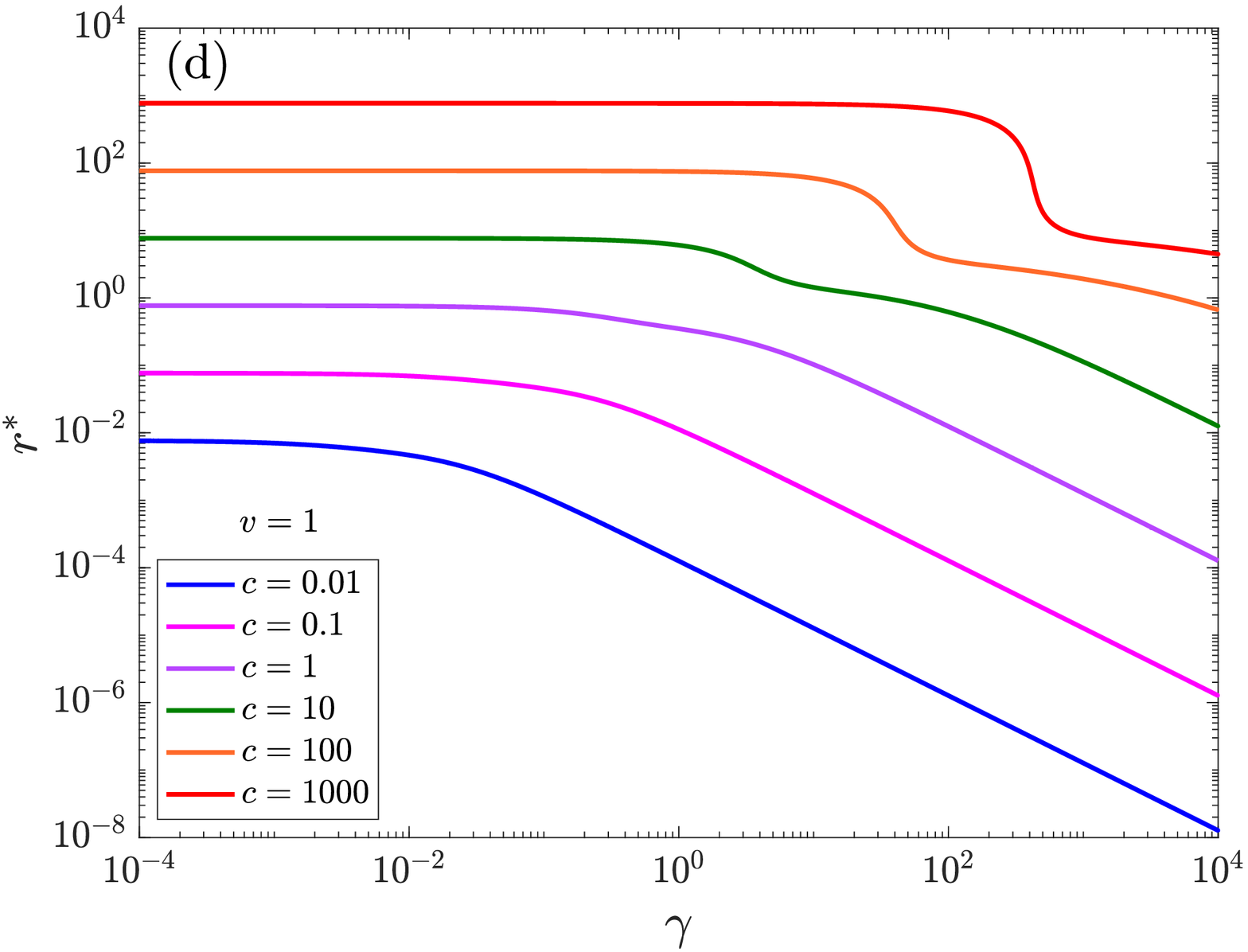}%
	}
	\caption{(a) Mean first-passage time for the Telegraphic process as a function of the resetting rate, with returns at constant speed and a target placed at position $ b=1 $. (b) The corresponding mean first-passage time in the ballistic regime and the optimal resetting rate (vertical dotted line), independent of $ v $. Data are obtained by simulating $ N=10^5 $ processes, with time step $ \mathrm{d}t=0.01 $ and $ \gamma=c=1 $. The theoretical curves (solid lines) are given by Eq. \eqref{eq:T_vconst} and the black lines represent instantaneous resetting. (c)-(d) Optimal resetting rate as a function of the system parameters, as obtained numerically, with $ b=1 $. The dotted horizontal line in panel (c) represents $ r^* $ in the case $ \gamma=0 $.}
	\label{fig:T_vconst}
\end{figure*}

In this case the return time is
\begin{equation}\label{key}
	\theta(x)=\frac{|x|}{v},
\end{equation}
where $ v $ is the speed. By inserting $ \theta(x) $ in Eq. \eqref{eq:phi_LT_surv} and using the expression of the survival probability, Eq. \eqref{eq:q_LT}, we compute the distribution of the duration of an unsuccessful subprocess in Laplace space, $ \hat{\varphi}(s) $, see Appendix \ref{app:surv_noninst}. The distribution $ \hat{\varpi}(s) $ can be evaluated immediately from Eqs. \eqref{eq:varpi_vs_f_LT} and \eqref{eq:f:_Tel_LT}, from which we obtain $ \hat{\varpi}(0) $:
\begin{equation}\label{eq:varpi_zero}
	\hat{\varpi}(0)=\frac{e^{-\tfrac bc\sqrt{r(r+2\gamma)}}}{R},
\end{equation}
where
\begin{equation}\label{key}
	R=1+\sqrt{\frac r{r+2\gamma}}.
\end{equation}
We can compute the first derivative as well, obtaining:
\begin{equation}\label{eq:varp_zero_prime}
	\hat{\varpi}'(0)=\left[\frac{R}{2}-1-\frac bc\left(r+\gamma\right)\right]\frac{e^{-\tfrac bc\sqrt{r(r+2\gamma)}}}{R\sqrt{r(r+2\gamma)}}.
\end{equation}
From these results and Eq. \eqref{eq:mean_FPT} it follows that the mean first-passage time is:
\begin{multline}\label{eq:T_vconst}
	\langle T\rangle=\frac 1r\left(Re^w-1\right)+\\\frac bv\bigg[\frac{2R\sinh w}{w}-
	1+\left(R-1\right)\left(\frac{2e^{-w}-1}{w}\right)\bigg],
\end{multline}
where $ w $ is the dimensionless variable
\begin{equation}\label{key}
	w=\frac{b}{c}\sqrt{r(r+2\gamma)}.
\end{equation}
We observe that in the limit $ v\to\infty $, the second term, which represents the contribution of the return phase, vanishes and we recover the result for instantaneous resetting \cite{EvaMaj-2018}. One can verify that for small $ r $ and fixed $ \gamma$, the mean first-passage time reads
\begin{equation}\label{key}
	\langle T\rangle\sim\frac{1}{\sqrt{2\gamma r}}\left(1+\frac{2\gamma b}{c}\right)+\frac bc+\frac{c}{\gamma v}\left(1+\frac{2\gamma b}{c}\right),
\end{equation}
where the last term is the contribution of the return motion. Such a contribution can be neglected for $ v\gg c\sqrt{2\gamma r}/\gamma $, and $ \langle T\rangle $ diverges as $ 1/\sqrt{r} $. On the opposite regime, i.e., for large $ r $, fixed $ v $ and $ \gamma $, the second term is of the same order as the first term and one obtains
\begin{equation}\label{key}
	\langle T\rangle\sim 2\left(1+\frac{c}{v}\right)\frac{e^{\frac{br}{c}}}{r},
\end{equation}
hence to leading order the mean first-passage time diverges as in the case of instantaneous resetting, but with a different coefficient.

It is useful to evaluate Eq. \eqref{eq:T_vconst} in the diffusive and ballistic limits. The first one is recovered for $ R\to 1 $, which yields
\begin{equation}\label{key}
	\langle T\rangle=\frac 1r\left(e^w-1\right)+\frac bc\left[\frac{2\sinh w}{w}-1\right],
\end{equation}
where in this case $ w $ is defined as
\begin{equation}\label{eq:w_diffusive}
	w=b\sqrt{\frac rD}.
\end{equation}
This is indeed the result obtained in Ref. \cite{BodSok-2020-BrResI}. It also interesting to check the opposite limit, which is recovered for $ R\to 2 $. One gets
\begin{equation}\label{eq:T_vconst_ball}
	\langle T\rangle=\frac 1r\left(2e^w-1\right)\left(1+\frac cv\right)-\frac bv,
\end{equation}
with
\begin{equation}\label{eq:w_ballistic}
	w=\frac{br}{c}.
\end{equation}
In Fig. \ref{fig:T_vconst} we represent the mean first-passage time of the process, described by Eq. \eqref{eq:T_vconst}, for a few values of the return speed. We also present the same observable in the ballistic regime, expressed by Eq. \eqref{eq:T_vconst_ball}. Both are compared with numerical simulations, showing good agreement with the theoretical behavior for all the considered values of the resetting rate. Remarkably, in the ballistic regime the optimal resetting rate $ r^* $ does not depend on $ v $. Indeed, one can verify that Eq. \eqref{eq:T_vconst_ball} yields the following equation for $ r^* $
\begin{equation}\label{key}
	e^{-z}=2-2z,\quad z=\frac{br}{c}
\end{equation}
whose solution $ z^* $ can be obtained in terms of the Lambert $ W $ function \cite{Olv}:
\begin{equation}\label{key}
	z^*=1+W\left(-\frac{1}{2e}\right)\approx 0.76803.
\end{equation}
This is confirmed by the behavior of the optimal resetting rate as a function of $ v $ and $ \gamma $, panels (c)  and (d). As $ \gamma $ decreases, the curves representing $ r^*$ become less dependent on $ v $ (left panel) and converge to a constant value (right panel) given by $ r^*=cz^*/b $.

\subsection{Return motion at constant acceleration}
\begin{figure*}
	\subfloat{
		\includegraphics[width=8.6cm]{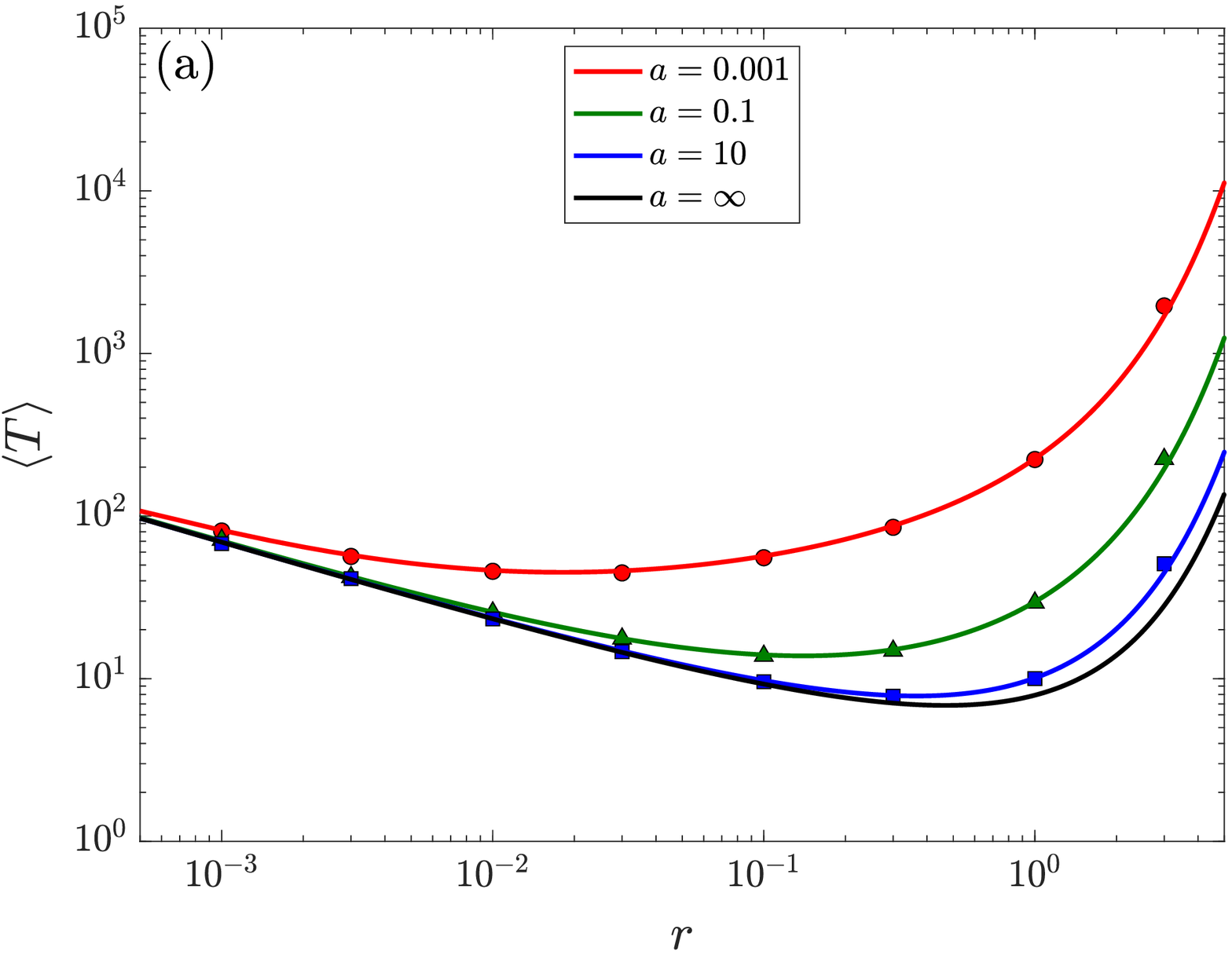}%
	}\quad
	\subfloat{
		\includegraphics[width=8.6cm]{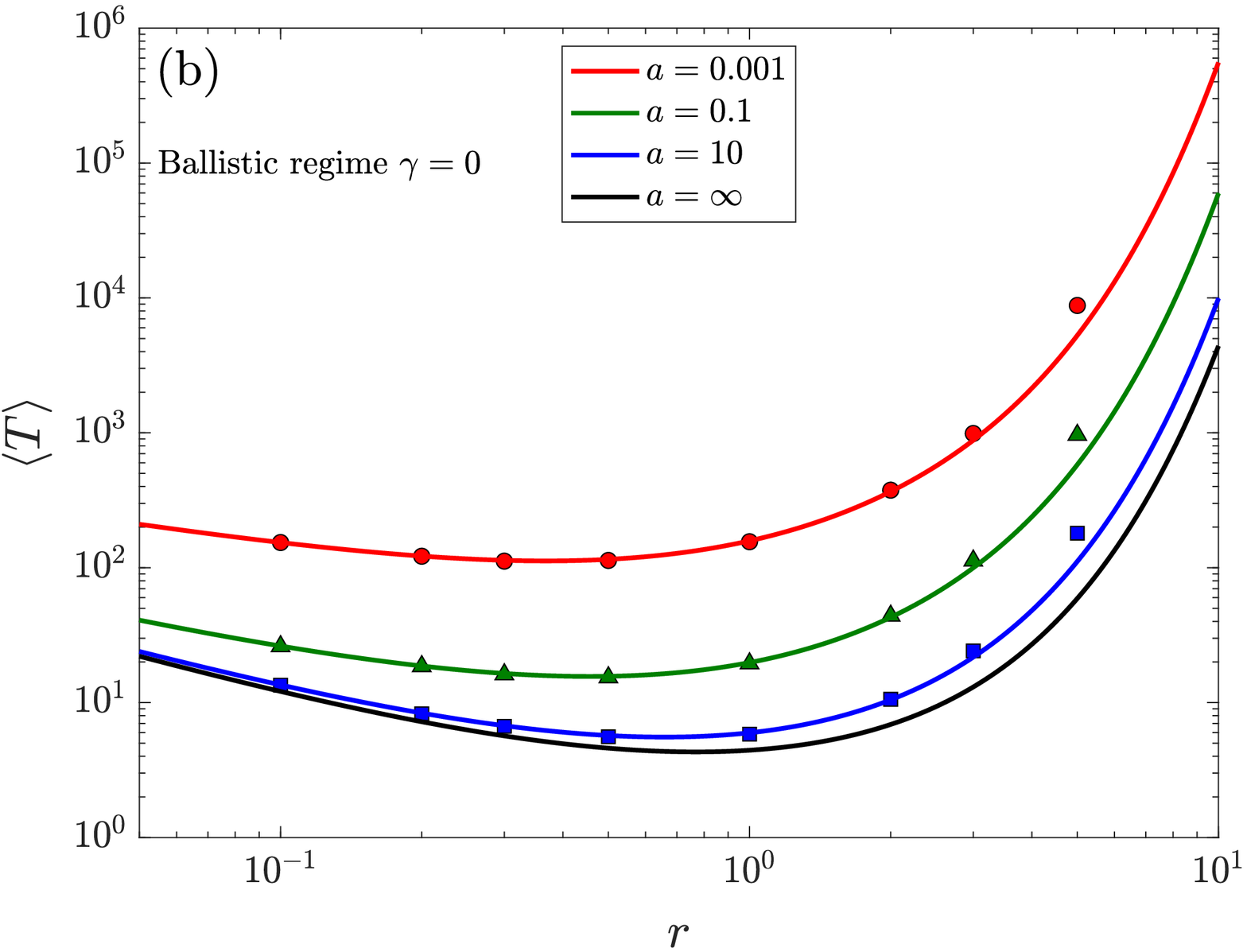}%
	}\hfill
	\subfloat{
		\includegraphics[width=8.6cm]{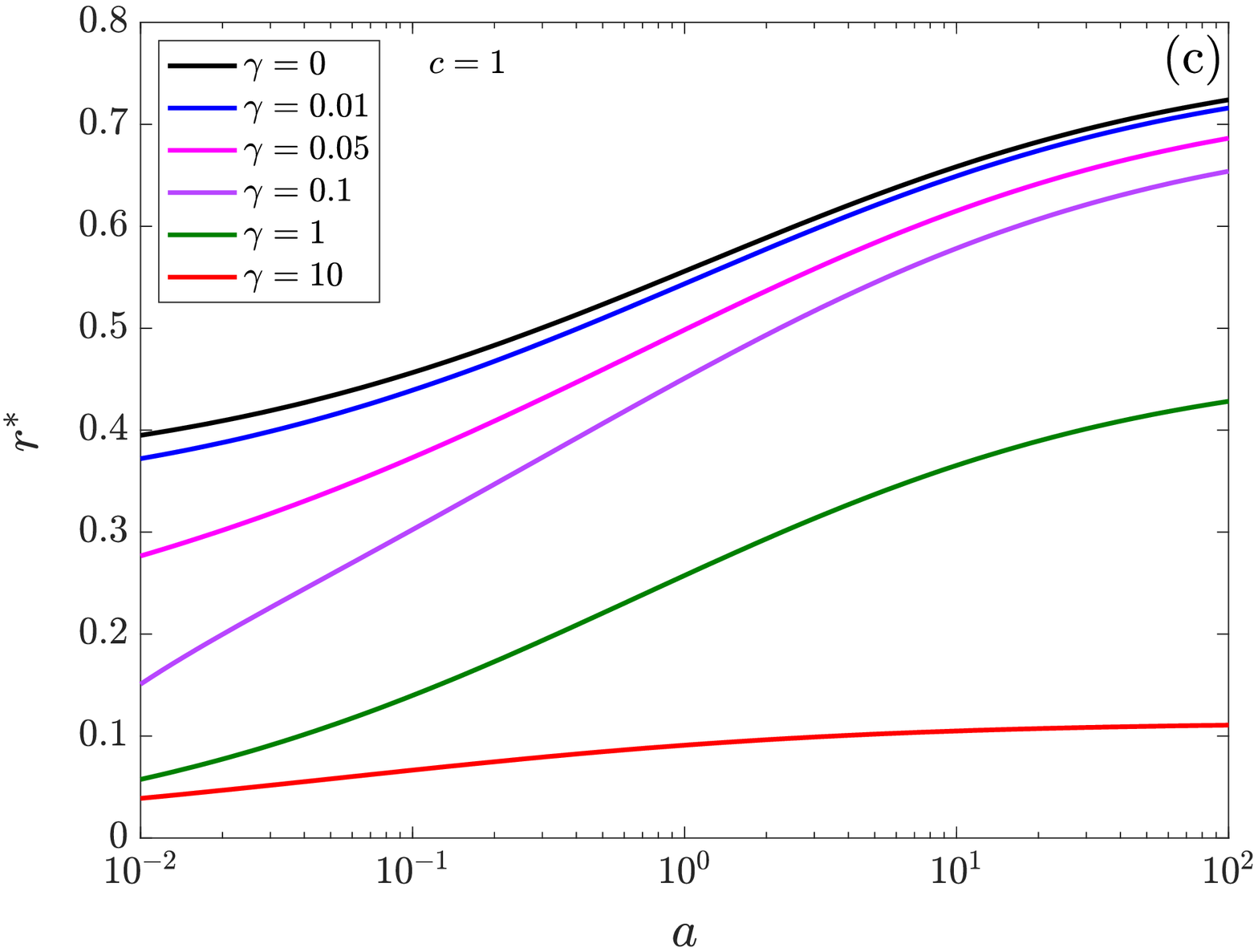}%
	}\quad
	\subfloat{
		\includegraphics[width=8.6cm]{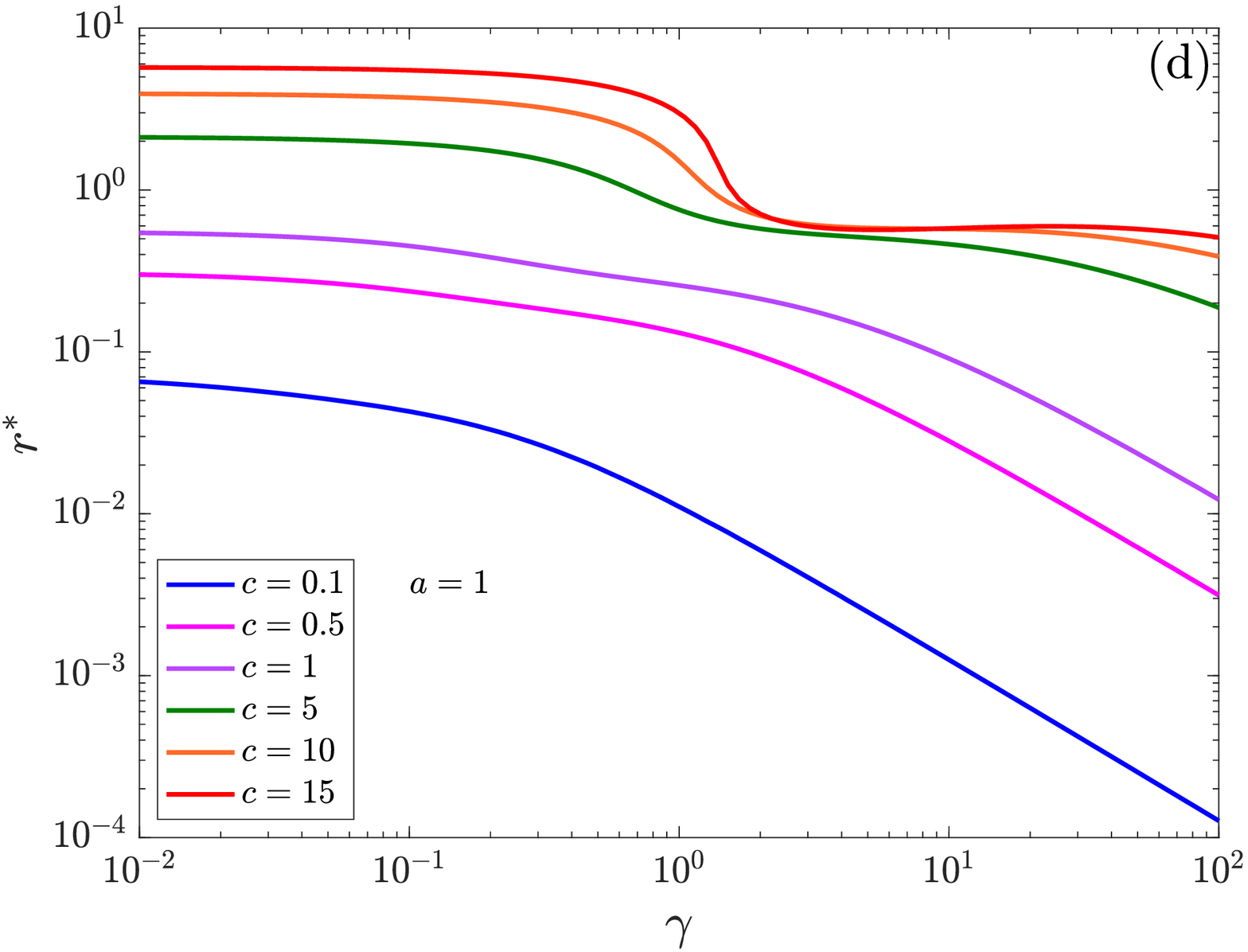}%
	}
	\caption{(a) Mean first-passage time for the Telegraphic process as a function of the resetting rate, with returns at constant acceleration and a target placed at position $ b=1 $. (b) The corresponding mean first-passage time in the ballistic regime. Data are obtained by simulating $ N=10^5 $ processes, with time step $ \mathrm{d}t=0.01 $ and $ \gamma=c=1 $. The theoretical curves (solid lines) are given by Eq. \eqref{eq:T_vconst} and the black lines represent instantaneous resetting. (c)-(d) Optimal resetting rate as a function of the system parameters, as obtained numerically, with $ b=1 $.}
	\label{fig:T_aconst}
\end{figure*}
The return time is given by
\begin{equation}\label{key}
	\theta(x)=\sqrt{\frac{2|x|}{a}},
\end{equation}
where $ a $ is the absolute value of the acceleration. The Laplace transform $ \hat{\varphi}(s) $ can be then computed, yielding the expression reported in Appendix \ref{app:surv_noninst}. Equations \eqref{eq:varpi_zero} and \eqref{eq:varp_zero_prime} yield the value of $ \hat{\varpi}(0) $ and $ \hat{\varpi}'(0) $, respectively, hence we obtain the following mean first-passage time:
\begin{multline}\label{eq:T_aconst}
	\langle T\rangle = \frac 1r\left(Re^w-1\right)-\sqrt{\frac{2b}{a}}+
	\sqrt{\frac{\pi b}{8aw}}\bigg\lbrace2R\cosh w+\\
	Re^w\mathrm{erf}\left(\sqrt{w}\right)+
	\left(2-R\right)e^{-w}\mathrm{erfi}\left(\sqrt{w}\right)-2e^{-w}\bigg\rbrace,
\end{multline}
where $ \mathrm{erfi}(z) $ is the imaginary error function, defined by
\begin{equation}\label{key}
	\mathrm{erfi}(z)=-i\mathrm{erf}(iz)=\frac{2}{\sqrt{\pi}}\int_{0}^{z}e^{t^2}\mathrm{d}t.
\end{equation}
Also in this case, we can recover the result of instantaneous resetting by taking the limit $ a\to\infty $. For small $ r $ and fixed $ \gamma $ we obtain
\begin{multline}\label{key}
		\langle T\rangle\sim\frac{1}{\sqrt{2\gamma r}}\left(1+\frac{2\gamma b}{c}\right)+\frac bc+\\
		\left(2\gamma r\right)^{\frac 14}\sqrt{\frac{\pi c}{8a\gamma^2}}\left(1+\frac{2\gamma b}{c}\right),
\end{multline}
where the contribution of the return phase is represented by the last term. For finite $ a $, this term can be neglected when
\begin{equation}\label{key}
	a\gg \frac{\pi c}{2}\sqrt{\frac{r^3}{2\gamma}},
\end{equation}
and $ \langle T\rangle $ diverges as $ 1/\sqrt{r} $, as in the previous case. On the contrary, for large $ r $, fixed $ \gamma$ and $ a $, the second term in Eq. \eqref{eq:T_aconst} yields the leading-order contribution and one obtains
\begin{equation}\label{key}
	\langle T\rangle\sim \sqrt{\frac{\pi c}{8ar}}e^{\frac{br}{c}},
\end{equation}
meaning that the mean first-passage time grows faster with respect to that considered in the case of returns at constant speed. 

By evaluating Eq. \eqref{eq:T_aconst} in the diffusive limit we obtain
\begin{multline}
	\langle T\rangle = \frac 1r\left(e^w-1\right)-\sqrt{\frac{2b}{a}}+
\sqrt{\frac{\pi b}{8aw}}\bigg\lbrace 2\sinh w+\\
e^{w}\mathrm{erf}\left(\sqrt{w}\right)+e^{-w}\mathrm{erfi}\left(\sqrt{w}\right)\bigg\rbrace,
\end{multline}
where $ w $ is defined as in Eq. \eqref{eq:w_diffusive}. In the ballistic regime instead the resulting mean first passage time is
\begin{multline}\label{key}
	\langle T\rangle =\frac 1r\left(2e^w-1\right)-\sqrt{\frac{2b}{a}}+\\
	\sqrt{\frac{\pi b}{2aw}}\left[1+\mathrm{erf}\left(\sqrt{w}\right)\right]e^w,
\end{multline}
where in this case the definition of $ w $ is given by Eq. \eqref{eq:w_ballistic}. The theoretical values of $ \langle T\rangle $ in both the intermediate and ballistic regimes are depicted in Fig. \ref{fig:T_aconst}, panels (a)-(b), and compared with our numerical simulations, showing good agreement. Panels (c) and (d) display the value of the optimal resetting rate with respect to the system parameters, as obtained numerically. We note that in this case the optimal resetting rate retains its dependence on the parameter $ a $ of the return phase also in the ballistic limit, as depicted in panel (c).

\subsection{Harmonic return motion}
\begin{figure*}
	\subfloat{
		\includegraphics[width=8.6cm]{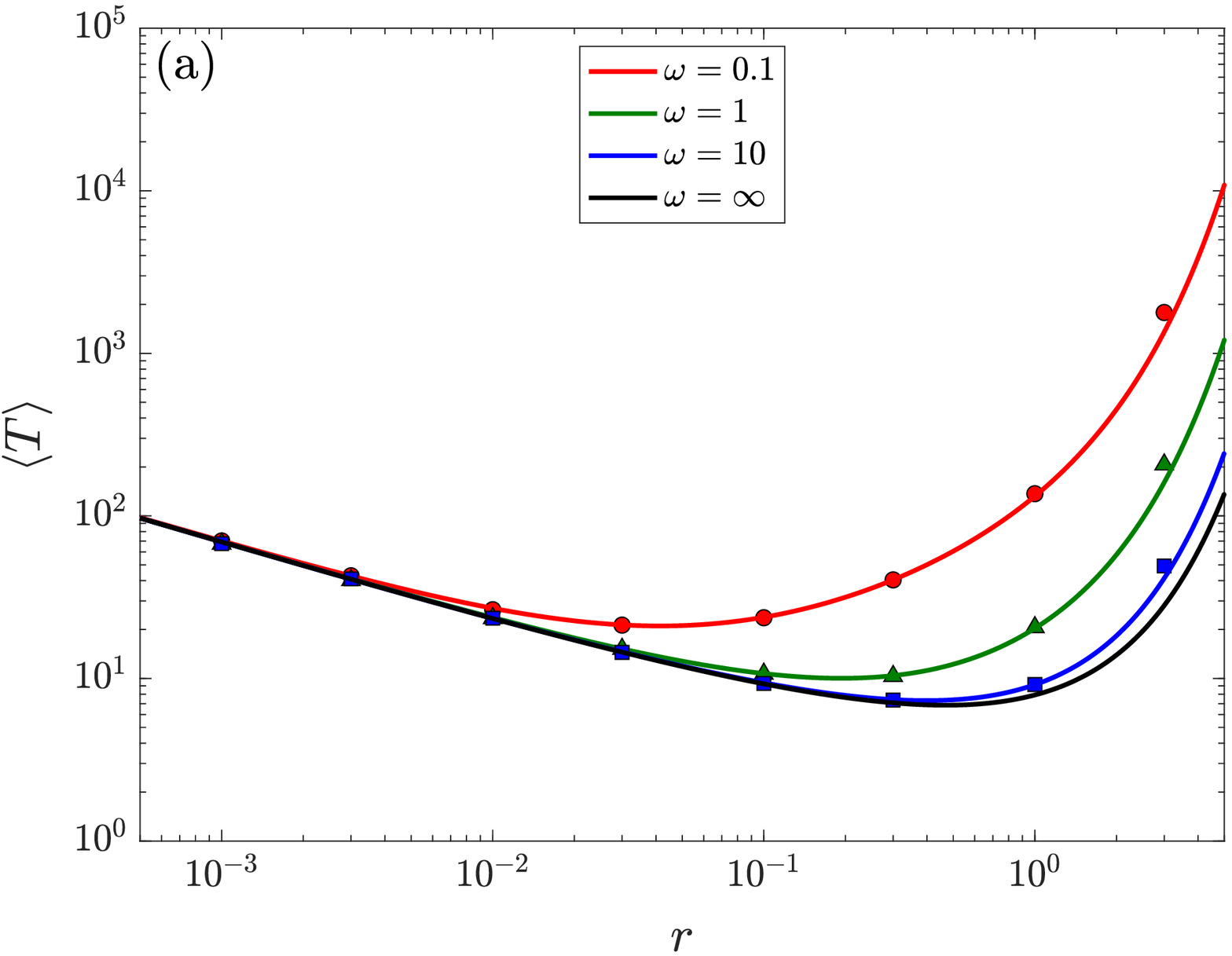}%
	}\quad
	\subfloat{
		\includegraphics[width=8.6cm]{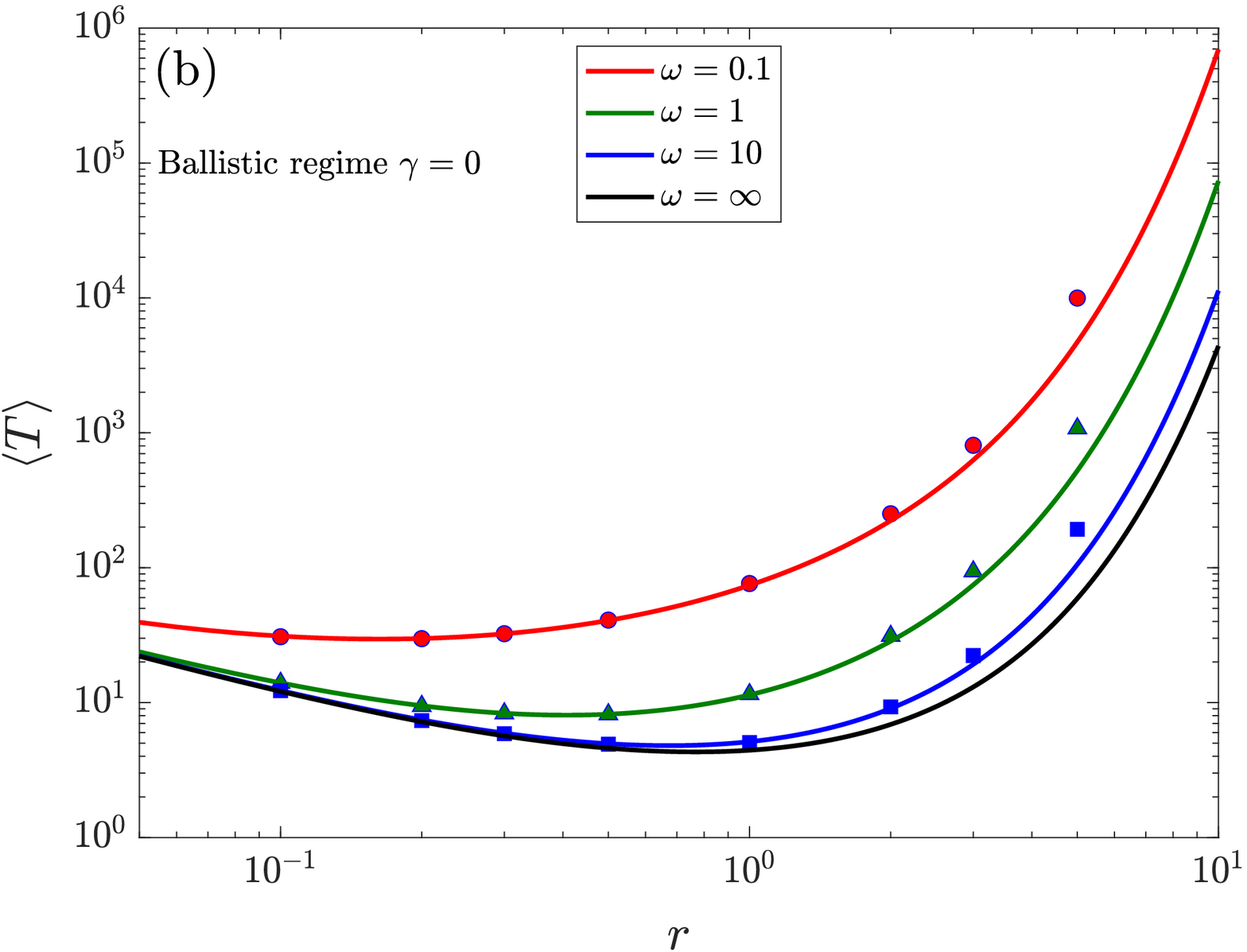}%
	}\hfill
	\subfloat{
		\includegraphics[width=8.6cm]{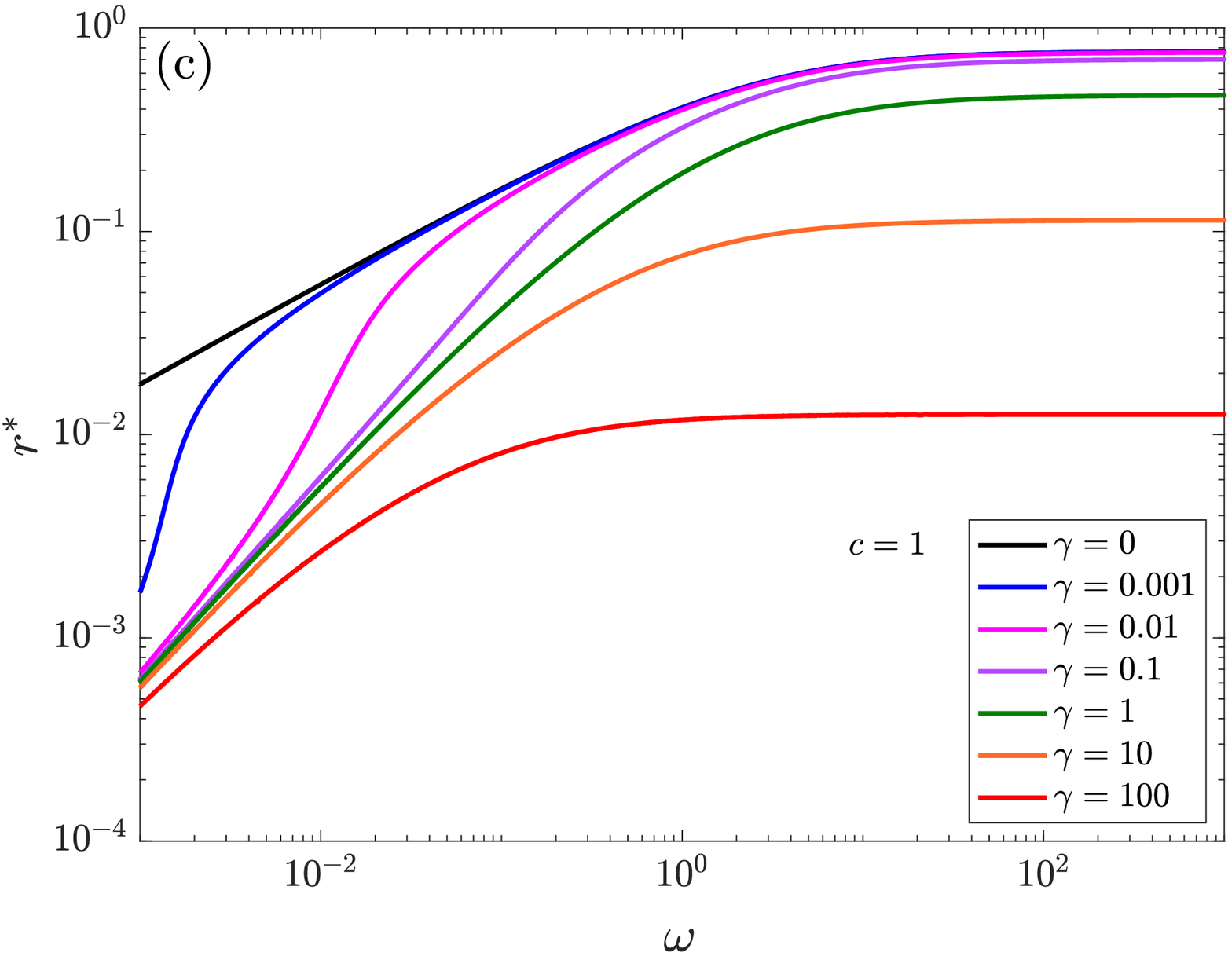}%
	}\quad
	\subfloat{
		\includegraphics[width=8.6cm]{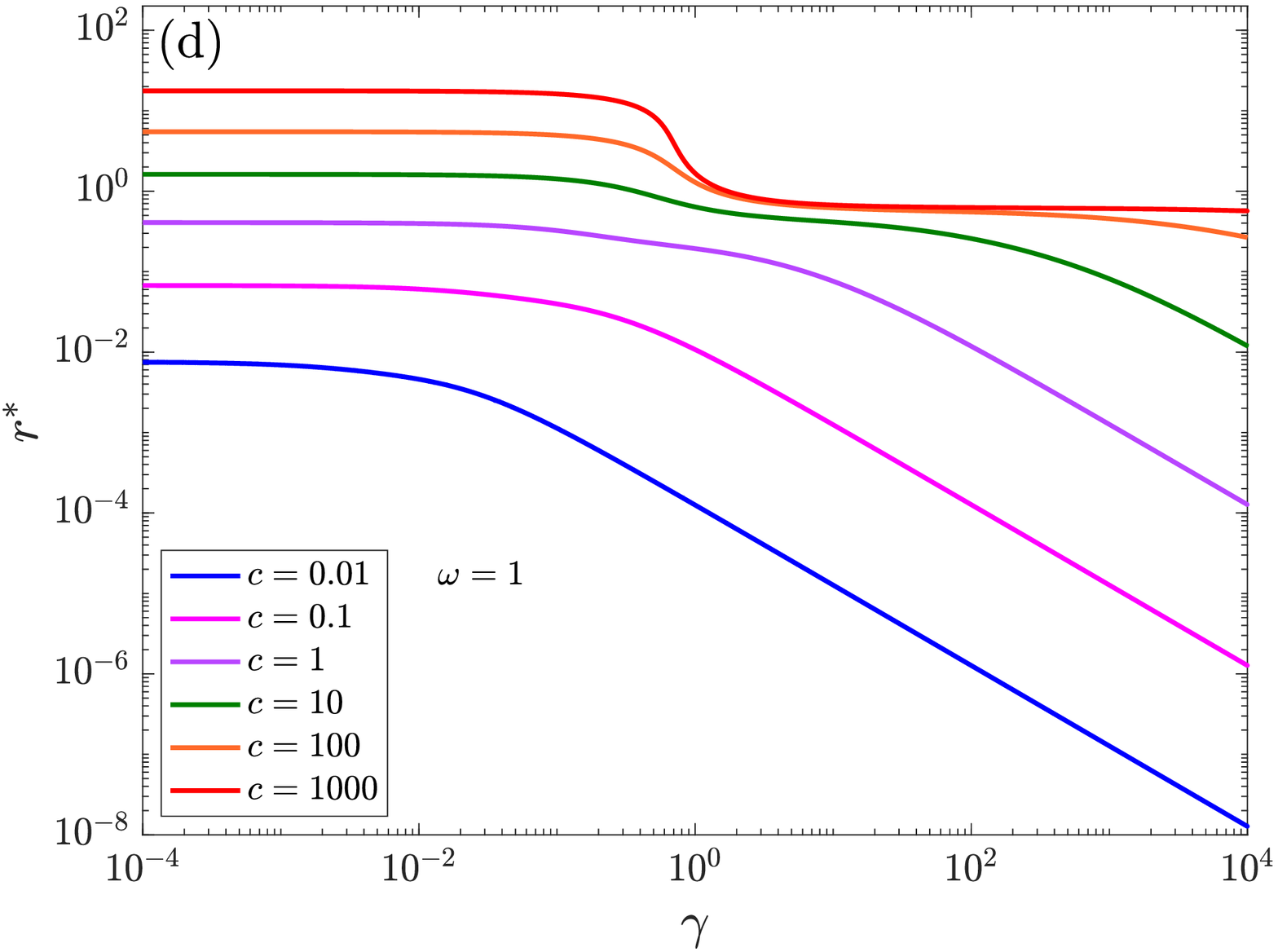}%
}
	\caption{(a) Mean first-passage time for the Telegraphic process as a function of the resetting rate, with returns under the effect of a harmonic potential and a target placed at position $ b=1 $. (b) The corresponding mean first-passage time in the ballistic regime. Data are obtained by simulating $ N=10^5 $ processes, with time step $ \mathrm{d}t=0.01 $ and $ \gamma=c=1 $. The theoretical curves (solid lines) are given by Eq. \eqref{eq:T_vconst} and the black lines represent instantaneous resetting. (c)-(d) Optimal resetting rate as a function of the system parameters, as obtained numerically, with $ b=1 $.}
	\label{fig:T_arm}
\end{figure*}
This case is almost trivial, because the return time is independent of the position at which the resetting occurs:
\begin{equation}\label{key}
	\theta=\frac{\pi}{2\omega}.
\end{equation}
From Eq. \eqref{eq:phi_LT_surv} it is quite easy to see that the computation of $ \hat{\varphi}(s) $ yields
\begin{equation}\label{key}
	\hat{\varphi}(s)=re^{-\tfrac{\pi s}{2\omega}}Q(b,s+r),
\end{equation}
see Appendix \ref{app:surv_noninst}, while $ \hat{\varpi}(0) $ and $ \hat{\varpi}'(0) $ are still defined as in the previous cases. A straightforward computation leads to
\begin{equation}\label{eq:T_arm}
	\langle T\rangle =\left(\frac 1r+\frac{\pi}{2\omega}\right)\left(Re^w-1\right),
\end{equation}
and the corresponding results for the diffusive and ballistic limit are simply recovered by setting $ R=1 $ or $ R=2 $ in the previous expression, and using the correct definition of $ w $, given by Eq. \eqref{eq:w_diffusive} and Eq. \eqref{eq:w_ballistic}, respectively. The limit $ \omega\to\infty $ yields the result of instantaneous resetting, but for finite $ \omega $ Eq. \eqref{eq:T_arm} suggests that one can neglect the effect of the return phase when $ \omega\gg \pi r/2 $. However, for large $ r $, fixed $ \omega $ and $ \gamma $, the mean first-passage time grows exponentially
\begin{equation}\label{key}
	\langle T\rangle\sim\frac{\pi}{\omega}e^{\frac{br}{c}},
\end{equation}
displaying the fastest growth of the three cases considered in this paper. This is due to the fact that, since the time cost to return is constant and independent of the resetting position, a higher number of resetting events implies a higher amount of time spent in performing the returns, noticeably increasing the mean first-passage time.
Figure \ref{fig:T_arm} displays the results of our simulations. In particular, panels (a) and (b) show the difference between the $ \gamma>0 $ and $ \gamma=0 $ regimes. In both cases, the numerical data show good agreement with the theoretical predictions. In panels (c) and (d) we show the behavior of the optimal resetting rate with the system parameters. As in the case of returns at constant acceleration, $ r^* $ is affected by the return phase also in the ballistic limit.

\section{Conclusions}\label{s:conc}
In this paper we have considered the one-dimensional Telegraphic process undergoing stochastic Poissonian resetting. Unlike the standard description of models with resetting, the return to the initial location is performed according to a deterministic law of motion, so that the time cost needed to perform the return is correlated with the position at which the resetting occurs. This leads to interesting consequences regarding the first-passage properties and the distribution of the process. In the present work, we have considered three kinds of return motion: (i) return at constant speed, (ii) constant acceleration and (iii) under the effect of a harmonic potential.

As it happens for Brownian motion, we have shown that Poissonian resetting stabilizes the whole process and for each type of return dynamics considered in this paper the system reaches a stationary state. However, the literature suggests that this may not hold true for different kinds of resetting protocols, for example, power-law waiting times between resetting events, in which case the process remains nonstationary in nature and hence possibly attracted in the domain of infinite ergodic theory rather than standard ergodic theory. Moreover, we have also shown that Poissonian resetting improves the first-passage properties of the system, in the sense that while the mean first hitting time of a target is infinite for the reset-free process, it becomes finite in the presence of resetting.

It is worth observing that the Telegraphic process displays both similarities and differences with respect to Brownian motion. The similarities are connected to the fact that there exists a scaling limit for the system parameters, namely the speed of propagation $ c $ and the reversal rate $ \gamma $, where the two processes are equivalent. This explains, e.g., the fact that the stationary distributions for Brownian motion and the Telegraphic process display the same functional form, see the results in \cite{BodSok-2020-BrResI}. The distributions are in both cases described by a parameter $ \lambda_r $, representing the typical resetting length - on average, a particle travels a distance $ \ell=1/\lambda_r $ before being reset. For Brownian motion, $ \ell_{\mathrm{BM}}=\sqrt{D/r} $, hence the typical distance is affected by the stochastic dynamics exclusively via the diffusion coefficient $ D $. For the Telegraphic process instead, $ \ell_{\mathrm{TP}}=c/\sqrt{r(r+2\gamma)} $, therefore in this case the typical distance is controlled by two different parameters of the stochastic phase. Note that one can set $ \ell_{\mathrm{BM}}=\ell_{\mathrm{TP}} $ by satisfying the equality $ D=c^2/(r+2\gamma) $, therefore in principle the Telegraphic process under Poissonian resetting with rate $ r $ can be seen as Brownian motion with a rate-dependent diffusion coefficient, in the sense that the two system reach the same stationary state. However, one still observes relevant differences due to finite-time effects, such as the presence of ballistic peaks representing the contribution of ballistic motion. Moreover, even though the two systems show the same dynamical transition in the temporal relaxation to the steady state, the large deviation functions have different expressions, compare with the results in \cite{MajSabSch-2015}. Other differences are due to the constraint of finite speed of propagation. For example, in the case of Brownian motion the typical length decreases for large resetting rates as $ 1/\sqrt{r} $, while in the case of the Telegraphic process it decreases faster, as $ 1/r $, and depends explicitly on the speed $ c $.

Regarding the effects of the return phase, returns at constant speed and acceleration display, in a sense, an opposite behavior with respect to returns performed under the action of a harmonic potential. Indeed, in the former case the probability density function and the steady state reached by the process are independent of the return dynamics, so that one obtains the same result of the standard situation with instantaneous resetting; in the latter case instead, the return law provides important contributions to the resulting stationary state, which is indeed evidently different from the aforementioned result. Furthermore, when one investigates the first-passage properties, it becomes clear that the return phase modifies the mean first-passage time in a highly nontrivial way. The corresponding contribution may be neglected only for very small values of the resetting rate, or by taking very large values of the parameters of the return dynamics (with respect to the resetting rate). 

We think that our work may be useful for those situations where it has been recognized that the Telegraphic process provides a more suitable description with respect to Brownian motion, e.g., the run-and-tumble dynamics of Escherichia coli bacteria, in which some sort of resetting mechanism must be taken into account and for which the time cost to return can not be neglected.

\begin{acknowledgments}
The author acknowledge financial support from PRIN Research Project No. 2017S35EHN ``Regular and stochastic behavior in dynamical systems'' of the Italian Ministry of Education, University and Research (MIUR).
\end{acknowledgments}

\appendix
\section{Normalization of the probability density function}\label{app:norm}
In order to prove the normalization of the PDF describing the complete process, it is sufficient to show that
\begin{equation}\label{eq:app_G_norm}
	\int_{-\infty}^{+\infty}\hat{G}(x,s)\mathrm{d}x=\frac{1-\hat{\phi}(s)}{s},
\end{equation}
where $ \hat{\phi}(s) $ is the Laplace transform of the subprocess duration. Indeed, the normalization condition in Laplace space reads
\begin{equation}\label{key}
	\int_{-\infty}^{+\infty}\hat{P}(x,s)\mathrm{d}x=\frac 1s,
\end{equation}
hence the condition of Eq. \eqref{eq:app_G_norm} follows from Eq. \eqref{eq:P_complete_LT} in the main text.

We first consider $ \hat{G}_1(x,s) $. We have:
\begin{align}\label{key}
	\int_{-\infty}^{+\infty}\hat{G}_1(x,s)\mathrm{d}x&=\int_{-\infty}^{+\infty}\mathrm{d}x\int_{0}^{\infty}\mathrm{d}te^{-st}\Psi(t)p(x,t)\\
	&=\int_{0}^{\infty}e^{-st}\Psi(t)\mathrm{d}t\\
	&=\hat{\Psi}(s),
\end{align}
where we changed the order of integration between $ x $ and $ t $ and used the normalization of $ p(x,t) $. Here $ \hat{\Psi}(s) $ is:
\begin{equation}\label{key}
	\hat{\Psi}(s)=\int_{0}^{\infty}\mathrm{d}te^{-st}\int_{t}^{\infty}\mathrm{d}t'\psi(t')=\frac{1-\hat{\psi}(s)}{s}.
\end{equation}

We now write $ \hat{G}_2(x,s) $ as in Eq. \eqref{eq:G2_interm} and integrate over $ x $, obtaining:
\begin{multline}\label{key}
	\int_{-\infty}^{+\infty}\hat{G}_2(x,s)\mathrm{d}x=\int_{0}^{\infty}\mathrm{d}t'e^{-st'}\psi(t')\times\\\int_{-\infty}^{+\infty}\mathrm{d}x_0p(x_0,t')\int_{0}^{\infty}\mathrm{d}ue^{-su}\Theta\left(\theta(x_0)-u\right).
\end{multline}
The integral in $ u $ yields
\begin{equation}\label{key}
	\int_{0}^{\infty}e^{-su}\Theta\left(\theta(x_0)-u\right)\mathrm{d}u=\frac 1s-\frac{e^{-s\theta(x_0)}}{s},
\end{equation}
therefore the remaining calculation can be seen as the sum of two contributions. The first contribution comes from the term $ 1/s $ and reads:
\begin{equation}\label{key}
	\frac 1s\int_{0}^{\infty}\mathrm{d}t'e^{-st'}\psi(t')\int_{-\infty}^{+\infty}\mathrm{d}x_0p(x_0,t')=\frac{\hat{\psi}(s)}{s}.
\end{equation}
The second contribution is written as
\begin{equation}\label{key}
	\frac 1s\int_{-\infty}^{+\infty}\mathrm{d}x_0e^{-s\theta(x_0)}\int_{0}^{\infty}\mathrm{d}t'e^{-st'}\psi(t')p(x_0,t'),
\end{equation}
which is, a part from the prefactor $ 1/s $, the definition of $ \hat{\phi}(s) $, see Eq. \eqref{eq:phi_LT} in the main text. Hence we obtain
\begin{equation}\label{key}
	\int_{-\infty}^{+\infty}\hat{G}_2(x,s)\mathrm{d}x=\frac{\hat{\psi}(s)-\hat{\phi}(s)}{s}.
\end{equation}
By adding the contributions of $ \hat{G}_1(x,s) $ and $ \hat{G}_2(x,s) $, we finally obtain Eq. \eqref{eq:app_G_norm}.

\section{Laplace transform of the subprocess duration}\label{app:mean_duration}
In order to evaluate the Laplace transform of the duration of a subprocess, Eq. \eqref{eq:phi_LT} in the main text, a preliminary computation requires the solution of the integral:
\begin{multline}\label{key}
	\mathcal{I}(x)=\frac r2\int_{0}^{\infty}e^{-(s+\gamma+r) \tau}\bigg\lbrace\delta(x-c\tau)+\delta(x+c\tau)+\\
	\frac{\gamma}{c}\left[I_0(z)+\frac{\gamma tI_1(z)}{z}\right]\Theta(c\tau-|x|)\bigg\rbrace\mathrm{d}\tau.
\end{multline}
The part containing the delta functions yields:
\begin{equation}\label{key}
	\frac r2\int_{0}^{\infty}e^{-p\tau}\left[\delta(x-c\tau)+
	\delta(x+c\tau)\right]\mathrm{d}\tau=\frac {r}{2c}e^{-\frac pc|x|},
\end{equation}
where $ p=s+r+\gamma $. The remaining part can be computed by first considering the change of variable $ y= c\tau $, so that one obtains
\begin{multline}\label{key}
	\frac{\gamma r}{2c^2}\int_{0}^{\infty}e^{-\frac{s+\gamma+r}{c}y}\bigg[I_0\left(\frac {\gamma}{c}\sqrt{y^2-x^2}\right)+\\
	\frac{ yI_1\left(\frac {\gamma}{c}\sqrt{y^2-x^2}\right)}{\sqrt{y^2-x^2}}\bigg]\Theta(y-|x|)\mathrm{d}y,
\end{multline}
which takes the form of a Laplace transform involving modified Bessel functions of the first kind. We make use of the explicit formulas \cite{Bat-I}
\begin{equation}\label{key}
	\int_{x}^{\infty}e^{-st}I_0\left(\alpha\sqrt{t^2-x^2}\right)\mathrm{d}t=\frac{e^{-x\sqrt{s^2-\alpha^2}}}{\sqrt{s^2-\alpha^2}},
\end{equation}
and
\begin{multline}
	\int_{x}^{\infty}e^{-st}\frac{t}{\sqrt{t^2-x^2}}I_1\left(\alpha\sqrt{t^2-x^2}\right)\mathrm{d}t=\\
	\frac{s}{\alpha\sqrt{s^2-\alpha^2}}e^{-x\sqrt{s^2-\alpha^2}}-\frac{e^{-xs}}{\alpha},
\end{multline}
both valid for $ x>0 $ and $ \mathfrak{R}(s)>\left\lvert\mathfrak{R}(\alpha)\right\rvert $; therefore, by considering also the first part, we arrive at:
\begin{equation}\label{}
	\mathcal{I}(x)=\frac{r}{2c}\sqrt{\frac{2\gamma+r+s}{r+s}}e^{-\frac{|x|}{c}\sqrt{(2\gamma+r+s)(r+s)}}.
\end{equation}
Hence, we get
\begin{equation}\label{key}
	\hat{\phi}(s)=\int_{-\infty}^{+\infty}e^{-s\theta(x)}\mathcal{I}(x)\mathrm{d}x,
\end{equation}
and more explicit expressions can be obtained once we specify the deterministic motion. In the case of returns at constant speed we have
\begin{equation}\label{key}
	\theta(x)=\frac{|x|}{v},
\end{equation}
so that the integration is straightforward and it is left to the reader. In the case of returns at constant acceleration we have
\begin{equation}\label{key}
	\theta(x)=\sqrt{\frac{2|x|}{a}},
\end{equation}
and we can use the formula
\begin{multline}\label{key}
		\int _{x_1}^{x_2}e^{-\alpha\sqrt{x}-\beta x}\mathrm{d}x = \frac 1\beta\bigg\lbrace e^{-\beta x_1-\alpha\sqrt{ x_1}}-\\
		e^{-\beta x_2-\alpha\sqrt{x_2}}-e^{\tfrac{\alpha^2}{4\beta}}\sqrt{\frac{\pi\alpha^2}{4\beta}}\bigg[\mathrm{erf}\left(\sqrt{\beta x_2}+\frac{\alpha}{2\sqrt{\beta}}\right)-\\
		\mathrm{erf}\left(\sqrt{\beta x_1}+\frac{\alpha}{2\sqrt{\beta}}\right)\bigg]\bigg\rbrace,
\end{multline}
where $ \alpha $, $ \beta $, $ x_1 $ and $ x_2 $ are non-negative parameters and $ \mathrm{erf}(z) $ is the error function:
\begin{equation}\label{key}
	\mathrm{erf}(z)=\frac{2}{\sqrt{\pi}}\int_{0}^{z}e^{-t^2}\mathrm{d}t.
\end{equation}
By applying this formula with $ \alpha=s\sqrt{2/a} $, $ \beta=\lambda_{r+s} $, $ x_1=0 $ and taking the limit $ x_2\to\infty $ we obtain
\begin{equation}\label{key}
	\hat{\phi}(s)=\frac{r}{r+s}\left\{1-\sqrt{\pi}\xi e^{\xi^2}\left[1-\mathrm{erf}\left(\xi\right)\right]\right\},
\end{equation}
where
\begin{equation}\label{key}
	\xi=\sqrt{\frac{s^2}{2a\lambda_{r+s}}}.
\end{equation}
Finally, in the case of harmonic motion, the time cost to return is independent of the starting position
\begin{equation}\label{key}
	\theta=\frac{\pi}{2\omega},
\end{equation}
and one trivially gets:
\begin{equation}\label{key}
	\hat{\phi}(s)=\frac{r}{r+s}e^{-\tfrac{\pi s}{2\omega}}.
\end{equation}

\section{Large deviation form of the PDF for the Telegraphic process under stochastic resetting}\label{s:Relax}
We have shown in the main text that when the return motion is performed at constant velocity or acceleration, the resulting PDF is the same as the case of instantaneous returns,
\begin{equation}\label{eq:app:P}
	P(x,t)=e^{-rt}p(x,t)+r\int_{0}^{t}e^{-rt'}p(x,t')\mathrm{d}t',
\end{equation}
where $ p(x,t) $ is given by
\begin{multline}
	p(x,t)=\frac{e^{-\gamma t}}{2}\bigg\lbrace\delta(x-ct)+\delta(x+ct)+\\
	\frac{\gamma}{c}\left[I_0(z)+\frac{\gamma tI_1(z)}{z}\right]\Theta(ct-|x|)\bigg\rbrace.
\end{multline}
Here the variable $ z $ is
\begin{equation}\label{key}
	z=\frac{\gamma}{c}\sqrt{c^2t^2-x^2}.
\end{equation}
By using the relation $ I'_0(y)=I_1(y) $, we can rewrite the integral on the r.h.s. of Eq. \eqref{eq:app:P} as
\begin{multline}\label{key}
	\int_{0}^{t}e^{-rt'}p(x,t')\mathrm{d}t'=\frac{e^{-(r+\gamma)t}}{2c}I_0(z)\Theta(ct-|x|)+\\\frac{r+2\gamma}{2c}\int_{|x|/c}^{t}e^{-(r+\gamma)t'}I_0\left(\frac{\gamma}{c}\sqrt{c^2t'^2-x^2}\right)\mathrm{d}t',
\end{multline}
therefore Eq. \eqref{eq:app:P} may be rewritten as:
\begin{multline}\label{eq:app:Pt}
	P(x,t)=\frac{e^{-(r+\gamma) t}}{2}\bigg\lbrace\delta(x-ct)+\delta(x+ct)+\\
	\frac{1}{c}\left[(r+\gamma)I_0(z)+\frac{\gamma^2 tI_1(z)}{z}\right]\Theta(ct-|x|)\bigg\rbrace+\\
	\frac{r(r+2\gamma)}{2c}\int_{|x|/c}^{t}e^{-(r+\gamma)t'}I_0\left(\frac{\gamma}{c}\sqrt{c^2t'^2-x^2}\right)\mathrm{d}t'.
\end{multline}
We observe that the PDF can be split in a singular part and a regular part. The singular part is that containing the delta functions, namely
\begin{equation}\label{key}
	P_{\mathrm{sing}}(x,t)=\frac{e^{-(r+\gamma)t}}{2}\left[\delta(x+ct)+\delta(x-ct)\right],
\end{equation}
which describes the contribution of ballistic motion, i.e., those walks that do not experience any velocity reversal up to time $ t $. Note that for the resetting-free process, the probability of this contribution decays as $ \exp(-\gamma t) $, while in this case the decay rate is increased by the positive parameter $ r $, and $ P\propto\exp\left[-(r+\gamma)t\right] $. The regular part instead describes the contribution of the diffusive part of $ P(x,t) $, which can be written in large deviation form. By using the asymptotic expansion of the Bessel functions and considering the change of variable $ t'=ut $ in the integral on the r.h.s. of Eq. \eqref{eq:app:Pt}, we may rewrite for large $ t $
\begin{multline}\label{key}
	P_{\mathrm{reg}}(x,t)\sim r(r+2\gamma)\sqrt{\frac{t}{8\pi\gamma c^2}}\int_{|x|/ct}^{1}\frac{\mathrm{d}u}{\sqrt{u}}e^{-tH(u,\frac{x}{ct})}+\\
	\frac{r+2\gamma}{\sqrt{8\pi\gamma c^2t}}e^{-tH\left(1,\frac{x}{ct}\right)},
\end{multline}
where
\begin{equation}\label{key}
	H(u,y)=(r+\gamma)u-\gamma u\sqrt{1-\frac{y^2}{u^2}}.
\end{equation}
For fixed $ y $, the function $ H(u,y) $ has a single minimum at
\begin{equation}\label{key}
	u_0=\frac{|y|}{\alpha}=\frac{r+\gamma}{\sqrt{r(r+2\gamma)}}|y|,
\end{equation}
which occurs within the integration limits as long as $ u_0<1 $. Hence for $ u_0<1 $ the integral can be estimated by the saddle-point method and one obtains
\begin{equation}\label{key}
	P_{\mathrm{reg}}(x,t)\sim e^{-\frac{|x|}{c}\sqrt{r(r+2\gamma)}}.
\end{equation}
On the other hand, if $ u_0>1 $ the main contribution to the integral is given by the region around $ u=1 $, thus the estimated term is of the same order as the first term, i.e.,
\begin{equation}\label{key}
	P_{\mathrm{reg}}(x,t)\sim e^{-t\left(r+\gamma-\gamma\sqrt{1-\frac{x^2}{c^2t^2}}\right)}.
\end{equation}
Therefore, the regular part of the PDF has the large deviation form $ P_{\mathrm{reg}}(x,t)\sim\exp\left[-tI(x,t)\right] $, with
\begin{equation}\label{key}
	I(x,t)=\begin{dcases}
		\frac{|x|}{ct}\sqrt{r(r+2\gamma)}&\text{for }|x|<\alpha ct\\
		r+\gamma-\gamma\sqrt{1-\frac{x^2}{c^2t^2}}&\text{for }|x|>\alpha ct.
	\end{dcases}
\end{equation}

\section{Survival PDF for the Telegraphic process}\label{app:surv}
It is convenient to begin with the discrete description of the Telegraphic process provided by the model of the persistent random walk \cite{Wei-2002}. Let us consider a particle moving with nearest-neighbor jumps on a discrete lattice. Each jumps is performed in a time $ \delta t $  with speed $ c $, so that the lattice spacing is $ \delta x=c\delta t $; suppose that at each step there is a probability $ \mathfrak{r} $ of reversing the direction of motion and a probability $ \mathfrak{t}=1-\mathfrak{r} $ of jumping in the same direction of the previous step. Call $ R(x,t) $ the probability of being at position $ x=i\delta x $, $ i\in\mathbb{Z} $, at time $ t=n\delta t $, $ n\in\mathbb{N} $,  with momentum directed to the right, and $ L(x,t) $ the corresponding probability for momenta directed to the left. Then $ R(x,t) $ and $ L(x,t) $ obey the following evolution equations \cite{ACOR}:
\begin{align}
	R(x,t+\delta t)&=\mathfrak{t}R(x-c\delta t,t)+\mathfrak{r}L(x+c\delta t,t)\label{eq:app_R}\\
	L(x,t+\delta t)&=\mathfrak{t}L(x+c\delta t,t)+\mathfrak{r}R(x-c\delta t,t).\label{eq:app_L}
\end{align}
Now suppose that an absorbing barrier is placed at $ b>0 $, so that no particle can arrive from $ b+c\delta t $. By evaluating the dynamics at the boundary, the previous system reduces to
\begin{align}
	R(b,t+\delta t)&=\mathfrak{t}R(b-c\delta t,t)\\
	L(b,t+\delta t)&=\mathfrak{r}R(b-c\delta t,t).
\end{align}
The evolution equations of the Telegraphic process can be recovered from Eqs. \eqref{eq:app_R} and \eqref{eq:app_L} by performing the continuum limit. We consider infinitesimal time steps, $ \delta t\to dt $, set
\begin{align}\label{key}
	R(x,t)&\approx r(x,t)dx\\
	L(x,t)&\approx l(x,t)dx,
\end{align}
and scale the transmission probability as \cite{Wei-2002}
\begin{equation}\label{key}
	\mathfrak{t}=1-\gamma dt,
\end{equation}
which corresponds to Poissonian statistics with rate $ \gamma $ for the velocity reversals. These choices lead to the system of equations
\begin{align}
	r(x,t+dt)&=\left(1-\gamma dt\right)r(x-cdt,t)+\gamma dt\,l(x+cdt,t)\nonumber\\
	l(x,t+dt)&=\left(1-\gamma dt\right)l(x+cdt,t)+\gamma dt\,r(x-cdt,t)\nonumber,
\end{align}
accompanied by the following dynamics at the boundary:
\begin{align}\label{key}
	r(b,t+dt)&=\left(1-\gamma dt\right)r(b-cdt,t)\\
	l(b,t+dt)&=\gamma dt\,r(b-cdt,t)\label{eq:app_l_boundary}.
\end{align}
Equation \eqref{eq:app_l_boundary} imposes that $ l(b,t) $ vanishes in the limit $ dt\to 0 $, hence by Taylor expanding the evolution equations for $ r(x,t) $ and $ l(x,t) $ we obtain in such a limit the following system of partial differential equations:
\begin{align}
	\frac{\partial r}{\partial t}&=-c\frac{\partial r}{\partial x}-\gamma\left(r-l\right)\label{eq:app_r}\\
	\frac{\partial l}{\partial t}&=c\frac{\partial l}{\partial x}+\gamma\left(r-l\right),\label{eq:app_l}
\end{align}
subject to the boundary condition
\begin{equation}\label{key}
	l(b,t)=0.
\end{equation}

In order to solve the problem, we introduce the functions $ q=r+l $ and $ w=r-l $, and write the corresponding equations which are obtained by summing and subtracting \eqref{eq:app_r} and \eqref{eq:app_l}:
\begin{align}
	\frac{\partial q}{\partial t}&=-c\frac{\partial w}{\partial x}\label{eq:app_q}\\
	\frac{\partial w}{\partial t}&=-c\frac{\partial q}{\partial x}-2\gamma\,w.\label{eq:app_w}
\end{align}
The function $ q(x,t) $ just defined is the survival PDF. The boundary condition on $ l(x,t) $ corresponds to a boundary condition on $ q(x,t) $. Note that since $ l(x,t) $ vanishes at the boundary, it follows from their definition that $ q(b,t) $ and $ w(b,t) $ are equal. Therefore, by evaluating Eq. \eqref{eq:app_w} at $ x=b $ and putting $ w(b,t)=q(b,t) $ we obtain the boundary condition for $ q(x,t) $:
\begin{equation}\label{key}
	2\gamma\,q(b,t)+\frac{\partial q(b,t)}{\partial t}+c\left.\frac{\partial q(x,t)}{\partial x}\right\rvert_{x=b}=0.
\end{equation}
One can show that the system of coupled equations \eqref{eq:app_q} and \eqref{eq:app_w} can be transformed in a system of decoupled Telegrapher's equations \cite{Wei-2002}. Hence the function $ q(x,t) $ is the solution of the problem:
\begin{empheq}[left=\empheqlbrace]{align}\label{key}
	&\frac{\partial^2 q}{\partial t^2}+2\gamma\frac{\partial q}{\partial t}=c^2\frac{\partial^2 q}{\partial x^2}\\
	&q(x,0)=\delta(x)\\
	&\left.\frac{\partial q(x,t)}{\partial t}\right\rvert_{t=0}=0\\
	&2\gamma\,q(b,t)+\frac{\partial q(b,t)}{\partial t}+c\left.\frac{\partial q(x,t)}{\partial x}\right\rvert_{x=b}=0.
\end{empheq}
This may be solved in Laplace space, by seeking solutions $ \hat{q}(x,s) $ which are linear combinations of the free problem and moreover satisfy the boundary condition. Hence we put
\begin{equation}\label{key}
	\hat{q}(x,s)=\hat{p}(x,s)+A(s)e^{-\lambda_s(2b-x)},
\end{equation}
where
\begin{equation}\label{key}
	\hat{p}(x,s)=\frac{\lambda_s}{2s}e^{-|x|\lambda_s},\quad\lambda_s=\frac{\sqrt{s(s+2\gamma)}}{c},
\end{equation}
is the solution of the free problem, while the other term does not yield contributions in the interval $ (-\infty,b] $ for $ t<b/c $. The boundary condition in Laplace space is
\begin{equation}\label{key}
	\left(s+2\gamma\right) \hat{q}(b,s)+c\left.\frac{\partial \hat{q}(x,s)}{\partial x}\right\rvert_{x=b}=0,
\end{equation}
hence by plugging our ansatz into this equation we get the expression for the unknown coefficient $ A(s) $:
\begin{equation}\label{key}
	A(s)=-\frac{\lambda_s}{2\gamma s}\left(s+\gamma-c\lambda_s\right).
\end{equation}
As a double check, we can compute the total survival probability we obtain from this expression. It is easy to see that
\begin{align}\label{key}
	Q(b,s)&=\int_{-\infty}^{b}\hat{q}(x,s)\mathrm{d}x\\
	&=\frac{1}{s}-\frac{1}{2\gamma s}\left(s+2\gamma-c\lambda_s\right)e^{-b\lambda_s},
\end{align}
which is indeed the survival probability already obtained in the literature \cite{MalJemKun-2018,EvaMaj-2018}.

\begin{widetext}
\section{Laplace transform of the duration of an unsuccessful subprocess }\label{app:surv_noninst}
The Laplace transform of the duration of an unsuccessful subprocess can be computed from Eq. \eqref{eq:phi_LT_surv} of the main text:
\begin{equation}\label{key}
	\hat{\varphi}(s)=r\int_{-\infty}^{b}e^{-s\theta(x)}\hat{q}(x,s+r)\mathrm{d}x,
\end{equation}
where
\begin{equation}\label{key}
	\hat{q}(x,s)=\frac{\lambda_s}{2s}\left[e^{-|x|\lambda_s}-\frac{1}{\gamma}\left(s+\gamma-c\lambda_s\right)e^{-\lambda_s(2b-x)}\right],
\end{equation}
and $ \theta(x) $ depends on the return motion. We recall that $ \lambda_s $ is defined as:
\begin{equation}\label{key}
	\lambda_s=\frac{\sqrt{s(s+2\gamma)}}{c}.
\end{equation}
In the following, it is convenient to define the variable
\begin{equation}\label{key}
	p=s+r.
\end{equation}
In the case of returns at constant speed $ v $, the return time is $ \theta(x)=|x|/v $ and thus we get:
\begin{equation}\label{key}
	\hat{\varphi}(s)=\frac{r\lambda_p/p}{s/v+\lambda_p}+\frac{r\left(\frac{p+2\gamma}{c^2}\right)\left(\frac{p+2\gamma-c\lambda_p}{2\gamma}\right)e^{-b(s/v+\lambda_p)}+\frac{rs\lambda_p}{vp}\left[\frac{p-c\lambda_p}{2\gamma}e^{-b(s/v+\lambda_p)}-\frac{p+\gamma-c\lambda_p}{\gamma}e^{-2b\lambda_p}\right]}{\left(s/v+\lambda_p\right)\left(s/v-\lambda_p\right)}.
\end{equation}
For the returns at constant acceleration $ a $, we have $ \theta(x)=\sqrt{2|x|/a} $. By defining the variables
\begin{align}\label{key}
	w &= b\lambda_p\\
	\xi&=\frac{s}{\sqrt{2a\lambda_p}},
\end{align}
and the function
\begin{equation}\label{key}
	F(z)=\frac{2}{\sqrt{\pi}}\int_{0}^{z}e^{t^2-2\xi t}\mathrm{d}t,
\end{equation}
we can write the Laplace transform as:
\begin{multline}
	\hat{\varphi}(s)=\frac rp\left(1-\frac{p+2\gamma-c\lambda_p}{2\gamma}e^{-w-2\xi\sqrt{w}}\right)-\frac{r\xi}{p}\sqrt{\frac \pi 4}\bigg\lbrace e^{\xi^2}\left[1+\mathrm{erf}\left(\sqrt{w}+\xi\right)-2\mathrm{erf}\left(\xi\right)\right]-\\
	\frac{p+\gamma-c\lambda_p}{\gamma}e^{\xi^2-2w}\left[F\left(\sqrt{\xi}\right)+\mathrm{erf}\left(\xi\right)-1\right]\bigg\rbrace.
\end{multline}
Finally, when the return motion is performed under the action of a harmonic potential, the return time is $ \theta(x)=\pi/2\omega $ and thus the Laplace transform is simply:
\begin{equation}\label{key}
	\hat{\varphi}(s)=re^{-\tfrac{\pi s}{2\omega}}\left[\frac 1p-\frac 1{2\gamma p}\left(p+2\gamma-c\lambda_p\right)e^{-b\lambda_p}\right].
\end{equation}
\end{widetext}

\section{Simulation method}
In order to perform the simulations of the Telegraphic process, we took advantage of Kac's representation
\begin{equation}\label{key}
	x(t)=c_0\int_{0}^{t}(-1)^{N(t')}\mathrm{d}t',
\end{equation}
where $ N(t) $ is the number of events up to time $ t $ of a homogeneous Poisson process of rate $ \gamma $. It follows that for small time steps $ dt $ we can write
\begin{equation}\label{key}
	x(t+dt)\approx x(t)+c_0 dt\cdot(-1)^{N(t)}.
\end{equation}
Instead of considering $ N(t) $, we can take instead a signal $ \sigma(t) $ switching between the values $ 0 $ and $ 1 $ with rate $ \gamma $. In other words, the probability that $ \sigma(t) $ changes state in the time interval $ (t,t+dt) $ is $ \gamma dt $, and the position in all our simulations was evolved according to:
\begin{equation}\label{key}
	x_{j+1}=x_j+c_0 dt\cdot(-1)^{\sigma(t)}.
\end{equation}

%

\end{document}